\documentclass[twocolumn,pra,aps,floatfix,amsmath,amssymb]{revtex4-2}
\usepackage{epsfig,graphicx,tabularx,epstopdf,bm}

\begin{document}

\title{dc atomtronic quantum interference device: quantum superposition of persistent-current states and a parity-protected qubit}

\author{H. M. Cataldo}

\affiliation{
Universidad de Buenos Aires, Facultad de Ciencias Exactas y Naturales,
Departamento de F\'isica. Buenos Aires, Argentina\\}
\affiliation{
Consejo Nacional de Investigaciones Cient\'ificas y T\'ecnicas - Universidad de Buenos Aires, Instituto de F\'isica de Buenos Aires. Buenos Aires, Argentina}

\begin{abstract}

A generalized Bose-Hubbard model in a two-mode approximation is applied to study the rotational dynamics of a direct-current atomtronic quantum interference device. Modified values of on-site interaction and pair-tunneling
parameters of the Hamiltonian, derived from the small-oscillation periods of the Josephson modes, are shown to
provide an excellent agreement to the Gross-Pitaevskii simulation results for the whole rotational frequency range,
reaching also the critical values of imbalance and current. This amounts to a full validation of the semiclassical
approximation of the modified Hamiltonian,
whose quantization is employed to investigate the quantum features of the stationary states.
Focusing on the frequency interval where the potential energy presents two minima, it is shown that the
central frequency, at which such minima are symmetric,
yields an atom number parity-protected qubit with a maximum entanglement of both persistent-current states,
 similar to those of superconducting circuits threaded by a half-quantum
of applied flux. Such a parity protection scheme survives within a small interval around
the central frequency, which sets the minimum rotational frequency precision that should be required to implement
the qubit. It is found that such a maximum admissible error in the frequency determination turns out to be inversely proportional to the qubit quality factor that measures the gap between the qubit energy levels and the following levels. 
It is shown that the chemical potential or condensate particle number
could be employed as suitable control parameters
to achieve the best trade-off between such qubit characteristics.
\end{abstract}
\maketitle

\section{Introduction}
Atomtronics has become a rapidly growing branch of research within the area of quantum science and technology,
which aims to manipulate ultracold atoms moving in matter wave circuits \cite{acircuits}.
In fact, cold atom quantum technologies realized to date permit coherent matter-wave manipulations with unprecedented control and precision over a wide range of physical configurations. 
Atomtronic devices sought to emulate known electronic components have been developed, like diodes \cite{pepino}, batteries \cite{zozu,caliga} and transistors \cite{caliga1}. Cold atom realizations of Josephson junctions (JJs) have been utilized to achieve the atomic counterpart of the celebrated superconducting quantum interference device (SQUID)
\cite{braginski},
known as the atomtronic quantum interference device (AQUID) \cite{acircuits}. Pioneering neutral-atom analogs of the SQUID had already begun to develop more than two decades ago, represented by superfluid helium quantum interference devices \cite{simmonds,satop}. 
Experimental realizations of AQUIDs started with a toroidal circuit of ultracold atoms interrupted by a weak link \cite{rama}. Rotation of the weak link gave rise to the atomtronic
counterpart of the radio frequency SQUID \cite{braginski}, 
which has been shown to generate well-defined phase slips between quantized persistent currents \cite{wright}, along with a phenomenon of winding number hysteresis and its accompanying fundamental excitations \cite{eckel}.  
On the other hand, the cold atom version of the dc SQUID \cite{braginski} was
obtained by diametrically establishing on a ring-shaped toroidal trap a couple of potential barriers that emulate
superconducting JJs \cite{boshier13}.
An imposed angular rotation of such barriers mimics the effect of the magnetic flux traversing the loop area of a SQUID
that leads to the quantum interference of currents referred to in the name of this device.
Such a phenomenon
occurs as well in the rotating AQUID, as was recently observed in the experiments of Ref.~\cite{ryu20}.
Actually, the current oscillations stemming from each JJ have identical amplitude, but they are generally out of phase, with a phase difference that depends on the rotation rate. Therefore, the amplitude of the resultant current varies as a result of such an interference. This phenomenon, observed in the experiment, must not be confused with the quantum superposition
(entanglement) of persistent-current states, which constitutes the key ingredient for an atomic qubit \cite{acircuits}. In
such a case, 
the basic engineering consists in breaking the rotational symmetry of a ring-shaped condensate by
inserting suitable weak-links, which open a gap between both persistent-current
states of opposite polarity at the degeneracy point. Thus, the symmetric and anti-symmetric combinations
of such states form the two states of the qubit \cite{solenov,hw,anun,anun0}.
Several qubit implementations of this kind have been proposed so far. In Ref. \cite{rf-aquid}, a ring-shaped condensate
with an additional lattice confinement interrupted by a single weak-link, was shown to be governed by an effective qubit
dynamics at degeneracy, 
where the two states of the qubit are the symmetric and anti-symmetric combinations of the clockwise and
anti-clockwise flow-states. Such a two-level effective dynamics was later demonstrated for an improved three
weak links architecture presenting a considerably enlarged parameter space \cite{3-wl}.
On the other hand, we are not aware of proposals for atomic qubits involving ring-shaped condensates
with an even number of potential barriers, such as the dc AQUID, that would be able to emulate certain superconducting circuit elements having an effective Josephson energy, which, in
contrast to the conventional JJs, turns out to be $\pi$-periodic in the phase difference across the element,
allowing only double Cooper pairs to tunnel.
Such basic elements can include two \cite{larsen20,smith20}, four \cite{glad}, or eight \cite{doucot} JJs, and
have been utilized as the main building blocks of several designs of protected superconducting qubits \cite{bell14,ioffe}.
Actually, each of such basic circuits yields a parity-protected qubit in which the two logical states are encoded by the parity of the number of Cooper pairs on a superconducting island. 

In this paper, we will analyze the conditions under which
a dc AQUID could behave as an atom number parity-protected qubit similar to the above superconducting circuit elements.
Taking into account the fundamental role played by the coherent tunneling of pairs of bosons (Cooper pairs)  in such
circuits, we will assume 
 a generalized Bose-Hubbard (GBH) model that includes pair-tunneling events. 
Although it has been shown that such a pair-tunneling
amplitude can be safely neglected for a non-rotating AQUID, even if a relative movement of the JJs of the kind carried out in the experiments \cite{boshier13,ryu20} is considered \cite{cat20}, 
we will see that for a significant range of rotation rates such an approximation turns out to be not valid.
Given the high particle numbers of the AQUIDs, the GBH Hamiltonian becomes a semiclassical Hamiltonian depending on
canonically conjugate variables given by the particle imbalance and phase difference between both halves of the AQUID.
Similarly, the Heisenberg equations become two-mode (TM) equations of motion for the canonical variables derived from the semiclassical Hamiltonian \cite{ragh99,anan06,jezek21}. 
A straightforward quantization of this Hamiltonian \cite{stringari} provides the theoretical framework to study the feasibility of a protected qubit. But, before this, a careful validation of the semiclassical Hamiltonian should be carried out by comparing the results arising from the TM equations of motion with those obtained from mean-field Gross-Pitaevskii (GP) simulations.
To meet such a requirement, we developed an alternative calculation of the pair-tunneling GBH parameter based on
the small oscillations around the GP energy minima, which produced excellent agreement up to the
critical values of imbalance and current for the entire range of rotation frequencies.
Once such a previous validation was successfully passed, we were able to investigate the quantum characteristics of the stable stationary states for the whole range of rotational frequencies, focusing especially on the vicinity of the frequency that yields a protected qubit.

This paper is organized as follows. In Sec.~\ref{sec2}, we describe
the Bose-Einstein condensates that form the AQUIDs analyzed in this work. Section~\ref{sec3} is
devoted to explain the GBH model in the TM approximation, along with the alternative calculation of the
pair-tunneling parameter, whose details can be found in Appendix~\ref{appA}.
Section~\ref{sec4} deals with the AQUID currents. The semiclassical persistent-current states are analyzed in
Sec.~\ref{sec4A}, while the critical imbalances and currents are discussed in Sec.~\ref{secrit}.
Section~\ref{sec6} contains the quantum treatment of the persistent-current states and discusses the feasibility of the
parity-protected qubit. Details concerning the solution of the Schr\"odinger equation of the quantum regime
can be found in Appendix~\ref{appB}.
Finally, the summary and conclusions of this work are gathered in Sec.~\ref{sec5}.

\section{The system and GP simulations}\label{sec2}
\begin{figure}
\includegraphics[width=\linewidth]{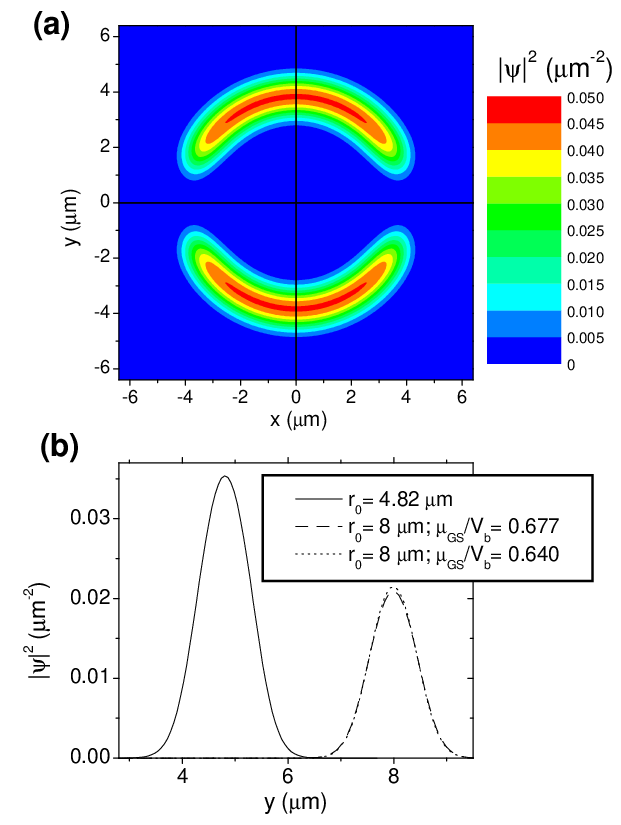}
\caption{(a): Contour density profile of the ground state of the non-rotating condensate with a ring radius $r_0=3.85\,\mu$m. (b): Bulk density profiles $|\Psi(x=0, y>0)|^2$ of the ground states of the non-rotating condensates 
with $r_0=4.82\,\mu$m (full line) and $r_0=8\,\mu$m, with $\mu_{\rm GS}/V_b=$ 0.677 (dashed line) and 
0.640 (dotted line).}
\label{figu1}
\end{figure}

We describe in what follows the condensates we have considered in our study, which were experimentally
realized as AQUIDs in Ref. \cite{ryu20}.
The trapping potential can be written as the sum of a term depending
on $x$ and $y$, and a term that is harmonic in the $z$ direction:
\begin{equation}
 V_{\rm ring}(r) + \frac{1}{2} m  \omega_z^2  z^2 \, ,
\label{vtrap}
\end{equation}
where  $r^2=x^2+y^2$ and  $m$ denotes the atomic mass of  $^{87}$Rb. 
The term depending on $r$ is modeled as the following  ring-Gaussian potential 
\begin{equation}
V_{\rm ring}(r)=    V_0 \, \left\{  1 - \exp\left[ - \frac{2}{w^2} (r-r_0 )^2\right]   \right\},
\label{toro}
\end{equation}
where $V_0$, $r_0$ and $w$ respectively  denote the depth,  radius and
$1/e^2$ width of the potential minimum.
The trap parameters have been selected according to those of Ref.
  \cite{ryu20}. We have set  $V_0/k_B$=82 nK, $w=1.7065\,\mu$m (corresponding to the  
 experimental radial trap frequency of 520 Hz),
 and three values of the radius
 $r_0$, namely 3.85, 4.82 and 8 $\mu$m, the first two of which were utilized in Ref. \cite{ryu20}. 
 The barrier potential reads,
\begin{equation}
V_{\rm barr}(y)=V_b\exp(-y^2/\lambda_b^2),
\label{barr}
\end{equation}
with the Gaussian height $V_b/k_B=42$ nK and width $\lambda_b=1.26118\,\mu$m, corresponding to the experimental full width at half maximum of 2.1 $\mu$m.
Under the conditions of the experiment \cite{ryu20},
a high value of the vertical trap frequency $\omega_z/(2\pi)=297$ Hz was assumed, which  yields
a quasi two-dimensional (2D) condensate allowing a simplified numerical treatment of the GP
simulations.
Thereby, the condensate order parameter is represented as a product
of a Gaussian wave function along the $z$ coordinate and a 2D 
order parameter normalized to one on the $x$-$y$ plane, $\Psi(x,y,t)$,
for which the corresponding GP equation in a rotating frame at the angular velocity $\Omega\,\hat{\bf z}$ 
reads \cite{castin,goldman},
\begin{equation}
[\hat{H}_0+gN|\Psi|^2]\Psi=i\hbar\frac{\partial \Psi}{\partial t},
\label{gp}
\end{equation}
with the non-interacting Hamiltonian,
\begin{equation}
\hat{H}_0=\frac{({\bf p}-{\bf A})^2}{2m}+V+W_{\rm rot},
\label{H0}
\end{equation}
where ${\bf p}$ denotes the momentum operator
 $-i\hbar(\hat{\bf x}\frac{\partial}{\partial x}+\hat{\bf y}\frac{\partial}{\partial y})$ and $V=V_{\rm ring}+V_{\rm barr}$.
The above Hamiltonian has the same form as that for a particle with a unit charge moving in a uniform magnetic
field ${\bf B}={\bf\nabla}\times {\bf A}=2m\Omega\,\hat{\bf z}$, with ${\bf A}=m\Omega(x\,\hat{\bf y}-y\,\hat{\bf x})$
the symmetric-gauge vector potential. Actually, artificial magnetic fields in ultracold gases can be engineered by a variety
of techniques ranging from the above simple rotation to laser-mediated angular momentum transfers \cite{goldman}.
The additional centrifugal potential $W_{\rm rot}=-m\Omega(x^2+y^2)/2$ in (\ref{H0})
repels the atoms away from the rotation axis $\hat{\bf z}$. The parameter
$N$ in the mean-field term of the GP equation (\ref{gp}) corresponds to the total number of particles, while $g$ denotes
the effective $2D$ coupling constant between the atoms \cite{castin},
\begin{equation}
g=g_{3D}\left(\frac{m\omega_z}{2\pi\hbar}\right)^{1/2},
\end{equation}
being $g_{3D}=4\pi\hbar^2a/m$ the three-dimensional coupling constant, with
$a= 98.98\, a_0 $ the  
$s$-wave scattering length of $^{87}$Rb and $a_0 $ the Bohr radius.
The particle number of each condensate was selected in order to yield values of the
chemical potential smaller than the barrier height $V_b$ (see Table \ref{tab1}). This
should ensure that the JJs are able to provide an adequate tunneling regime.
\begin{table}
\caption{ Characteristic parameters for each condensate; $\mu_{\rm GS}$ denotes the chemical
potential of the non-rotating ground state of the GP equation (\ref{gp}).}
\begin{ruledtabular}
\begin{tabular}{lcccc}
$r_0$ ($\mu$m)   & N &  $\mu_{\rm GS}/V_b$  &  $f_0=\Omega_0^{\rm (num)}/(2\pi)$ (Hz) & $\hbar/(2\pi m r_0^2)$ (Hz) \\
\hline
 3.85  & 3000 & 0.876 &   7.895  & 7.773\\
 4.82  & 2700 & 0.717 &   5.025 & 4.960 \\
 8.00  & 4500 & 0.677 &   1.8094  & 1.800\\
 8.00  & 4000 & 0.640 &   1.8094  & 1.800\\
\end{tabular}
\end{ruledtabular}
\label{tab1}
\end{table}
Figure \ref{figu1} shows density profiles of the non-rotating ground state of each condensate studied in this work,
which present healing lengths around 0.2 $\mu$m at their maximum densities. 
We have considered a single value of the chemical potential for each condensate radius, except for the highest radius
 $r_0=8\,\mu$m, for which we have considered a couple of values
yielding quite similar density profiles (Fig.~\ref{figu1}). However, it will be shown in Sec.~\ref{sec6}
that such configurations exhibit important differences in their energy spectra. Here it is worthwhile noticing  
that the rotating stationary states show fairly similar density profiles,
as well as chemical potential values, with respect to those of the non-rotating ground states.

There is an additional important characteristic feature of the rotating stationary states, which is the periodicity of
the circulating
current with respect to the imposed rotation rate of the condensate.
We may easily understand such a feature by considering the case of a one-dimensional (1D) ring of radius $r_0$, for which the rotating stationary state with the atoms circulating with a unit winding number yields a vanishing current in a rotating frame when
the imposed angular velocity reaches the value $\Omega_0=\hbar/(m r_0^2)$. Such a periodicity is well known for superconducting rings \cite{bloch}, where the current as a function of the magnetic flux through the ring has the period
of the magnetic flux quantum $\Phi_0=h/(2e)$, being $h$ Planck's constant and $e$ the electron charge. Here one has
a direct analogy with our superfluid ring, since the `magnetic flux' corresponding to $\Omega_0$ turns out to be
$2m\Omega_0\pi r_0^2=h$, which coincides with the value of the magnetic flux quantum, as
the unit charge of particles in Hamiltonian (\ref{H0}) should be identified with the Cooper pair charge $2e$.
We give in Table \ref{tab1} the numerically obtained values of the above period, $\Omega_0^{\rm (num)}$, which turn out to be
slightly higher than the 1D approximation $\Omega_0$. These results show decreasing percentage differences as the radius increases, indicating that the discrepancy simply arises from the dimensionality of the approximation.

Finally, we mention that the GP equation was numerically solved using the 
split-step Crank-Nicolson algorithm for imaginary- and real-time propagation
on a 2D spatial grid of 257$\times$257 points \cite{kumar19}.

\section{GBH model in the TM approximation}\label{sec3}
We will take as our starting point the generalized 
lowest-band Bose-Hubbard Hamiltonian \cite{duttar} arising from the 
many-body second-quantized Hamiltonian written in terms
of the TM approximation of the boson field operator,
\begin{equation}
\hat \Psi_{TM} (x,y)=\psi_u (x,y) \hat a_u + \psi_l (x,y) \hat a_l,
\label{fop}
\end{equation}
where $\psi_k (x,y)$ denotes the
wave function of a boson localized in the $k$-well 
with a corresponding annihilation operator denoted by $\hat a_k$
(here we denote $k=u$  and $l$ for the  upper and
lower well, respectively, in Fig.~\ref{figu1} (a)).
Such a Hamiltonian reads,
\begin{eqnarray}
\hat H_{BH} &=& 
- K (\hat a_u^{\dagger}\hat a_l + \hat a_l^{\dagger}\hat a_u) +
\frac{U}{4}(\hat a_u^{\dagger}\hat a_u -
\hat a_l^{\dagger}\hat a_l )^2\nonumber \\
&+& \frac{P}{2N}(\hat a_u^{\dagger}\hat a_u^{\dagger}\hat a_l\hat a_l + 
\hat a_l^{\dagger}\hat a_l^{\dagger}\hat a_u\hat a_u) \nonumber\\
&+& \frac{2P'}{N} \hat a_u^{\dagger}\hat a_u\hat a_l^{\dagger}\hat a_l,
\label{BH}
\end{eqnarray}
where we have replaced the operator $\hat a_u^{\dagger}\hat a_u+\hat a_l^{\dagger}\hat a_l$ by the total
number of particles $N$. The GBH model parameters read,
\begin{subequations}\label{param}
\begin{equation}
K= -\int \int dx\, dy\,\, \psi_{u}^* \left[\hat{H}_0
+\frac{gN}{2}\left(|\psi_u|^2+|\psi_l|^2\right)\right]  \psi_{l},\label{jota0}
\end{equation}
\begin{equation}
U= \frac{g}{2} \int \int dx\, dy\,\,  \left(|\psi_u|^4+|\psi_l|^4\right),\label{U0R}
\end{equation}
\begin{equation}
P= gN  \int \int dx\, dy\,\,   (\psi_{u}^*)^2 \,  \psi_{l}^2,\label{ijota}
\end{equation}
\begin{equation}
P'= gN  \int \int dx\, dy\,\,   |\psi_{u}|^2 \,  |\psi_{l}|^2 ,\label{ijotap}
\end{equation}
\end{subequations}
where all of them turn out to be real numbers \cite{jezek21} and
represent the following processes \cite{duttar},
\begin{subequations}
\begin{equation}
K\to \mbox{full single-particle tunneling}
\end{equation}
\begin{equation}
U\to \mbox{on-site interaction}
\end{equation}
\begin{equation}
P\to \mbox{pair tunneling}
\end{equation}
\begin{equation}
P'\to \mbox{neighbour interaction},
\end{equation}
\end{subequations}
here the full single-particle tunneling parameter $K$ embodies the contributions of
the conventional and density-induced single-particle tunneling processes \cite{duttar}.
The terms in $P$ and $P'$ are often neglected in standard Bose-Hubbard Hamiltonians. Here, although
both parameters turn out to be of the same order $P'\simeq |P|$, only the interaction term in the Hamiltonian
becomes negligible since we have $U\gg P'/N$. On the contrary, we will see that both tunneling parameters
$K$ and $P$ turn out to be comparable for certain values of the rotation frequency, and so the term 
proportional to $P$ in the
GBH Hamiltonian must be retained.

It is important to remark that the above localized states, characterized by the wave functions $\psi_u$
and $\psi_l$, and the corresponding creation and annihilation
operators, may actually depend on the number of particles at each well. 
Particularly,
as a most significant effect of the repulsive interparticle interaction, there is a broadening 
of the wave functions $\psi_{k}$ with increased occupation numbers \cite{cat11,duttar}. 
However, since the occupation number variations will keep small enough for our condensates,
we can safely disregard
such a dependence in the Hamiltonian (\ref{BH}).
We will assume a macroscopic occupation of states, which allows the replacement of creation and annihilation operators by complex $c$-numbers,
\begin{equation}
\hat a_k\to \sqrt{N_k}\, \exp(i\phi_k), 
\label{repl}
\end{equation}
where $\phi_k$  and $N_k$ represent the global phase \cite{jezek21,nigro20}
and particle number in
the $k$-well, respectively. Thus, one may define a phase difference between both wells as, 
\begin{equation}
\phi=\phi_u-\phi_l.
\end{equation}

There is a simple relationship between the single-particle tunneling parameter $K$ and
the energy-per-particle splitting between both
stationary solutions of the GP equation (\ref{gp}) yielding the lowest condensate energies \cite{jezek21},
\begin{equation}
E_{\pi}-E_0=2K,
\label{deltaE}
\end{equation}
where $E_0$ ($E_{\pi}$) denotes the energy per particle of the stationary state with $\phi=0$ ($\phi=\pm\pi$).
Actually the above localized wave functions are obtained from such stationary solutions as follows,
\begin{eqnarray}
\psi_u = \frac{1}{\sqrt{2}}(\Psi_0-\Psi_{\pi}) \\
\psi_l = \frac{1}{\sqrt{2}}(\Psi_0+\Psi_{\pi}),
\end{eqnarray}
where $\Psi_0$ ($\Psi_{\pi}$) denotes the 2D order parameter of the stationary state with 
$\phi=0$ ($\phi=\pm\pi$). Since they are orthogonal, $\int \int dx\, dy\,\Psi_0^*\Psi_{\pi}=0$,
the same occurs for the above wave functions $\psi_u$ and $\psi_l$, as expected. 
The stationary states can also be identified by their winding numbers.
In fact, within the rotational frequency interval $0<f<f_0$, the states with $\phi=0$ and $\phi=\pm\pi$ have winding
number 0 and 1, respectively. 
However, later we will see that the rotating condensate can have 
stationary states with any value of the phase difference $\phi$.  
So, we will prefer to identify the stationary states by the value of such a phase difference, calling them simply as 
$\phi$-states, and similarly for the eventual Josephson oscillations around such states, which will be 
referred to as $\phi$-modes.

The system dynamics is ruled by the Heisenberg equations
$d\hat a_k/dt=(i/\hbar)[\hat H_{BH},\hat a_k]$,
which under the replacement (\ref{repl}) leads to 
the following TM equations of motion, 
\begin{subequations}
\begin{equation}
 \hbar  \dot{Z}  = -2K 
\sqrt{1-Z^2}\,\sin\phi   +   P \,  (1 - Z^2) \sin (2 \phi) \label{zpun}
\end{equation}
\begin{equation}
\hbar  \dot{\phi}   = Z\left[  NU 
+ \frac{2K}{\sqrt{1-Z^2}}\cos\phi-P \cos (2 \phi)\right] , \label{fipun}
\end{equation}
\label{TMeqs}
\end{subequations}
where 
$Z=(N_l-N_u)/N$ denotes the particle imbalance between both wells.

The energy per particle
can be obtained in the TM approximation by making the replacement  (\ref{repl})
in the GBH Hamiltonian (\ref{BH}),
\begin{eqnarray}
E_{TM}(Z,\phi)&=& \frac{NU}{4}(1+Z^2)
- K\sqrt{1-Z^2}\cos \phi\nonumber\\
&+& \frac{P}{4}(1-Z^2)\cos 2\phi, \label{ener}
\end{eqnarray}
where we note that the difference $E_{TM}(0,\pm\pi)-E_{TM}(0,0)$ yields
the correct energy gap (\ref{deltaE}). Here it is worthwhile noticing that the expression (\ref{ener})
also arises from the 2D energy functional
\begin{equation}
\int \int dx\, dy\,\,  \Psi^* \left[\hat{H}_0
+\frac{1}{2}Ng|\Psi|^2\right]  \Psi
\label{Efun}
\end{equation}
 with the TM order parameter 
\begin{equation}
\Psi_{TM}(x,y)=\left[e^{i\phi}\psi_u(x,y)\sqrt{1-Z}+\psi_l(x,y)\sqrt{1+Z}\right]/\sqrt{2},
\label{psiTM}
\end{equation}
which stems from making the replacement (\ref{repl}) in Eq.~(\ref{fop}).
The equations of motion (\ref{TMeqs}) can also be 
written in the Hamiltonian form
\begin{equation}
\dot {\cal N} = -\frac{\partial {\cal H}}{\partial\phi}\,\, ; \,\, \dot\phi = \frac{\partial {\cal H}}{\partial {\cal N}},
\label{form}
\end{equation}
where ${\cal N}=NZ/2$ denotes the number of bosons that have tunneled (counted positively
from above to below in Fig.~\ref{figu1} (a)),
${\cal H}=NE_{TM}/\hbar $ corresponds to
 the semiclassical Hamiltonian and  (${\cal N}$,$\phi$) represents the canonically conjugate variables.

\begin{figure}
\includegraphics{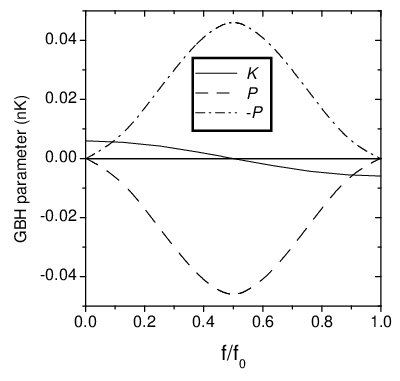}
\caption{Dependence on the rotational frequency of the GBH tunneling parameters $K$ and $\pm P$
for the condensate with the radius $r_0=3.85\,\mu$m. Similar graphs were obtained for the remaining
condensates with $r_0=4.82\,\mu$m and $8\,\mu$m.}
\label{figu2}
\end{figure}

It is instructive to analyze the energy (\ref{ener}) for small departures from the stationary values of $Z$
and $\phi$,
\begin{eqnarray}
E_{TM}(Z,\phi)-E_{TM}(0,0)&\simeq&  \frac{NU}{4}Z^2 \nonumber\\
&+& \frac{(K-P)}{2}\phi^2\label{so0}\\
E_{TM}(Z,\phi\pm\pi)-E_{TM}(0,\pm\pi)&\simeq&  \frac{NU}{4}Z^2 \nonumber\\
&-&\frac{(K+P)}{2}\phi^2\label{sopi},
\end{eqnarray}
where we have used in the coefficients of $Z^2$ the fact that
the parameters $K$ and $P$
turn out to be quite negligible with respect to $NU$ (see Table \ref{tab2} and Fig.~\ref{figu2}).
We have verified the accuracy of the above expressions as compared to the values obtained directly from
 the 2D energy functional (\ref{Efun}) with the corresponding TM wave function (\ref{psiTM}) with $Z^2\ll 1$.
According to Eq.~(\ref{U0R}), the coefficient of $Z^2$ in (\ref{so0}) and (\ref{sopi}) is positive definite, whereas,
taking into account Fig.~\ref{figu2}, we observe that
those of $\phi^2$ change their signs depending on the value of the 
rotation frequency. However, it is important to realize that the fact of
having a positive definite energy departure in (\ref{so0}) or (\ref{sopi}) does not constitute
a sufficient condition for the existence of a minimum in the 2D energy functional (\ref{Efun}). 
On the other hand, if the coefficient of  $\phi^2$ in (\ref{so0}) or (\ref{sopi}) turns out to be negative,
this ensures that the corresponding stationary state must correspond to an energy saddle. This is the case for the frequencies above the intersection of $K$ and $P$ ($f/f_0\simeq 0.9$) in Fig.~\ref{figu2}, where according to
Eq.~(\ref{so0}), the stationary states with $Z=0$ and $\phi=0$ must have an energy saddle. 
The same occurs for the 
rotational frequencies below the intersection of  $K$ and $-P$ ($f/f_0\simeq 0.1$) in Fig.~\ref{figu2},
where according to Eq.~(\ref{sopi}), the stationary states with $Z=0$ and $\phi=\pm\pi$ must correspond to energy saddles. However, later we will see that
such frequency intervals showing energy saddles actually turn out to be much wider than here predicted.

It has been shown in Refs. \cite{cat20,cap13,nigro17,jezek21}
that the agreement between the TM model results and the GP simulations turns out to be
substantially improved by
replacing the on-site interaction parameter $U$ by a lower effective value $U_{\rm eff}$, which
arise from the deformation that suffer the condensate densities at both wells
due to the departure of the particle imbalance from its vanishing stationary value.
Even though such a physical explanation for the discrepancy between $U$ and $U_{\rm eff}$
probably remains partially valid within the present situation, 
we have found that this is not the case for the formula
for $U_{\rm eff}$ utilized previously, since it now yields a clear overestimate.
So, a different procedure described in Appendix \ref{appA} was employed in this work to extract the value of $U_{\rm eff}$.
We show in Table \ref{tab2} the
values obtained for the different  condensates,
which lead to excellent agreements between the GP simulation and GBH model results.
\begin{table*}
\caption{GBH model parameters and width of the central interval in Fig.~\ref{figu3} 
for each condensate. The value of $P$ and the ratio $P_{\rm eff}/P$ correspond to the
frequency $f=f_0/2$.}
\begin{ruledtabular}
\begin{tabular}{lcccccc}
$r_0$ ($\mu$m) &   $\mu_{\rm GS}/V_b$   & $U$ (nK) & $P$ (nK) & $U_{\rm eff}/U$ &  $P_{\rm eff}/P$ &  $\Delta f/f_0$ \\
\hline
  3.85  & 0.876 & 0.01435 & -0.0462 & 0.8192 &  0.01549  & 0.07474\\
 4.82  & 0.717 & 0.01214 & -0.00335 & 0.8701 &  0.010274 & 0.02039 \\
 8.00  & 0.677 & 0.006821 & -0.00514 & 0.8964 &  0.002749 & 0.02358 \\
 8.00  & 0.640 & 0.006993 & -0.00214 & 0.9075 &  0.003012 & 0.01559 \\
\end{tabular}
\end{ruledtabular}
\label{tab2}
\end{table*}

\begin{figure}
\includegraphics{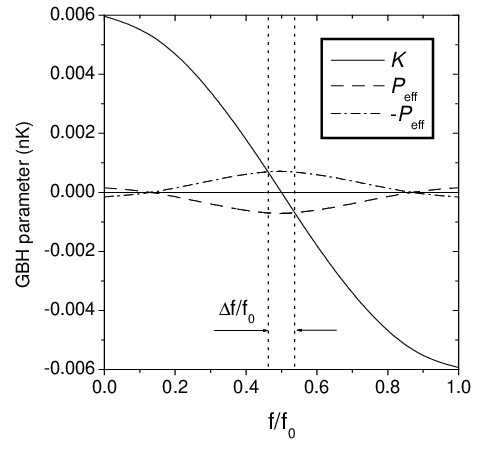}
\caption{Same as figure \ref{figu2} for the modified pair tunneling parameter $P_{\rm eff}$.
For rotation frequencies within the central interval 
limited by the vertical dotted lines, both states zero and $\pi$ have an energy minimum.}
\label{figu3}
\end{figure}

Even though this is the only correction to the GBH model parameters for non-rotating condensates
that is needed to restore the agreement with the GP simulation results, there arises in the rotating case an important additional discrepancy that again may be fixed by making a parameter correction. In fact, as explained in Appendix \ref{appA},
the pair tunneling coefficient $P$ given by Eq.~(\ref{ijota}) has to be replaced 
by a modified value $P_{\rm eff}$, which 
for most rotation rates turns out to be only a small fraction of the original value (see Fig.~\ref{figu3} 
and Table \ref{tab2}). 
One might consider such a 
reduction as somehow not unexpected, since a pair-tunneling amplitude much larger than that of single particles 
(Fig.~\ref{figu2}) seems to be difficult to interpret. In addition, such a reduction of the original value (\ref{ijota})
 could be physically interpreted as reflecting a quite softened effective repulsion between the pair of correlated atoms at tunneling. On the other hand, it has been shown that the tunneling in higher orbitals can have a large net effect
 on tunneling amplitudes. This suggests that a more formal treatment of the modified GBH parameters
 could be explored from multiorbital dressing procedures of the corresponding processes \cite{duttar}. 

As shown in Appendix \ref{appA}, the central frequency interval $\Delta f/f_0$ of Fig.~\ref{figu3}, which fulfills 
$|K/P_{\rm eff}|<1$, yields zero and $\pi$ states with energy minima, 
whereas at the left (right) of such an interval,
the state with $\phi=\pm\pi$ ($\phi=0$) corresponds to an energy saddle. 
According to Eq.~(\ref{deltaE}), at the center of such an interval
($f/f_0=1/2$), both energy minima have the same depth, while at the left (right) of such a central frequency
and within the interval, the 0-state should be stable (metastable) and the $\pi$-state should be metastable (stable).
One can find in Table \ref{tab2} a clear trend that seems to anticipate that in the limit of a dilute condensate
($\mu_{\rm GS}/V_b\ll 1$)
we will have $U_{\rm eff}\to U$ and $\Delta f/f_ 0\to 0$. On the other hand, the behavior of the
ratio $P_{\rm eff}/P$ seems to be difficult to interpret.

Finally, we will write the modified version of the TM equations of motion
(\ref{TMeqs}) that are obtained by replacing the parameters $U$ and $P$ by
their modified values $U_{\rm eff}$ and $P_{\rm eff}$, and also by
taking into account in Eq.~(\ref{fipun}) that
the contributions proportional to the single-particle tunneling $K$ and the pair tunneling
$P$ (or its effective value),
turn out to be quite negligible with respect to the term proportional to the on-site interaction 
$U$ (or its effective value). So, the modified TM equations read,
\begin{subequations}
\begin{equation}
 \hbar  \dot{Z}  = -2K
\sqrt{1-Z^2}\,\sin\phi   +   P_{\rm eff} \,  (1 - Z^2) \sin (2 \phi) \label{zp}
\end{equation}
\begin{equation}
\hbar  \dot{\phi} =  N U_{\rm eff} Z.
\label{fip}
\end{equation}
\label{TMeqsm}
\end{subequations}
From the above equations it is easy to verify that any state with $Z=0$ and 
$\phi=\pm\arccos(K/P_{\rm eff})$ should correspond to a stationary state.
Actually,
for frequencies within the central interval of Fig.~\ref{figu3},
the couple of saddles in between the energy minima at $\phi=\pm\pi$ and $\phi=0$
correspond to such $\pm\arccos(K/P_{\rm eff})$-states (see also the phase-space plot of Fig. 4 in Ref. \cite{jezek21}).

\section{AQUID currents}\label{sec4}
We will assume that the currents flowing across the JJs can be described from the rotating frame as,
\begin{equation}
I_k=I_0[\sin \varphi_k + \alpha_0 \sin 2\varphi_k],
\label{Ivsphase}
\end{equation}
where the label $k$ can take the values $l$ and $r$ denoting, respectively, the left and right JJ in Fig.~\ref{figu1} (a).
We will consider that a current $I_k$ circulating counterclockwise (i.e., the direction of the imposed rotation) 
has a positive sign,
and so the phase difference $\varphi_k$ across the JJ (phase jump) means `upper-phase minus lower-phase'
for the right JJ and vice versa for the left JJ.
A nonvanishing coefficient $\alpha_0$ in Eq.~(\ref{Ivsphase})
assumes a current-phase relationship that differs from the ideal sinusoidal one by the addition of a second-harmonic
contribution \cite{golubov,scazza}. The above coefficients $I_0$ and $\alpha_0$ depend only on the characteristics
of the JJ, but they are independent of the rotation rate of the system. The connection with the GBH model follows
by identifying the net current flowing from the upper to the lower portion of the condensate
 in Fig.~\ref{figu1} (a) as $N\dot Z/2=I_l-I_r$. So, 
with a little algebra we may write,
\begin{equation}
\hbar\dot Z = -4\hbar\frac{I_0}{N}(\cos\xi\, \sin\phi
+\alpha_0\cos 2\xi\, \sin 2\phi ),
\label{dotZ}
\end{equation}
with
\begin{equation}
 \xi  = \frac{\varphi_r+\varphi_l}{2}+n\pi
\label{xi}
\end{equation}
and
\begin{equation}
\phi =  \frac{\varphi_r-\varphi_l}{2}+n\pi,
\end{equation}
where $n$ denotes the winding number.
From Eq.~(\ref{dotZ}), taking into account (\ref{zp}) with $Z^2\ll 1$, we obtain 
the values of the current amplitude $I_0$ and the second-harmonic coefficient $\alpha_0$ in terms of the GBH parameters
of the non-rotating condensate $K=K^0$ and $P_{\rm eff}=P_{\rm eff}^0$,
\begin{eqnarray}
I_0  =  \frac{NK^0}{2\hbar}\\
\alpha_0  =  -\frac{P_{\rm eff}^0}{2K^0}.
\end{eqnarray}
We display in Table \ref{tab3}
 the values of such parameters for each condensate.
\begin{table}
\caption{ Current parameters for each condensate and maximum absolute value of the stationary currents.}
\begin{ruledtabular}
\begin{tabular}{lcccc}
$r_0$ ($\mu$m)   & $\mu_{\rm GS}/V_b$   &  $I_0/N$ (s$^{-1}$) &  $\alpha_0$ &  $\max(|I_k/N|)$ (s$^{-1}$) \\
\hline
 3.85  & 0.876 &  0.39067 &  -0.013321  & 0.39081\\
 4.82  & 0.717 &  0.07029 &   -0.0027587 & 0.07040 \\
 8.00  & 0.677 &  0.02524 &   -0.0017754 & 0.02550 \\
 8.00  & 0.640 &  0.01722 &   -0.0011648 & 0.01743 \\
\end{tabular}
\end{ruledtabular}
\label{tab3}
\end{table}

\subsection{Persistent-current states}\label{sec4A}
\begin{figure}
\includegraphics{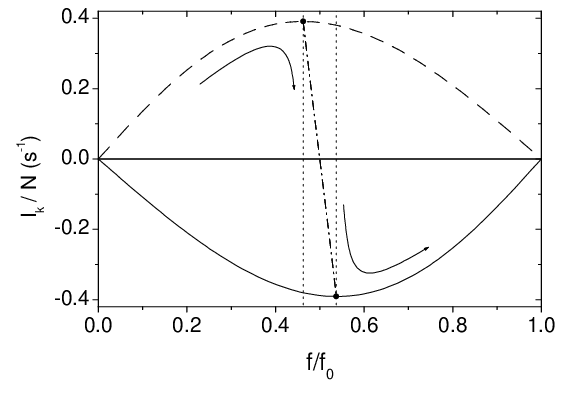}
\caption{Stationary currents $I_k$ in the condensate of radius $r_0=3.85\,\mu$m.
The full and dashed lines correspond to zero and $\pi$ states, respectively. The vertical dotted lines,
which limit the same central interval of Fig.~\ref{figu3}, mark the extremes of such currents indicated by
the circles.
The dash-dotted line corresponds to the saddle $\phi$-state ($\pi>\phi>0$), while the curved arrows represent
the saddle `trajectory' from $f=0$ to $f=f_0$.}
\label{figu5}
\end{figure}

We depict in Fig.~\ref{figu5} the
stationary currents of the condensate with radius $r_0=3.85\,\mu$m.
The 0- and $\pi$-states respectively present clockwise and counterclockwise currents in the frequency interval
$0<f<f_0$ of Fig.~\ref{figu5}. The extremes of such currents are easily obtained from Eq.~(\ref{Ivsphase})
for a phase jump fulfilling $\cos\varphi_k+2\alpha_0\cos 2\varphi_k=0$ which yields,
\begin{equation}
\varphi_k =\arccos\left (  -\frac{1}{8\alpha_0}-\sqrt{\frac{1}{2}+\frac{1}{64\alpha_0^2}}\right ).
\label{sol}
\end{equation}
We display in Table \ref{tab3} the maximum absolute value of such currents for each condensate.
As regards the current of the saddle $\phi$-state, it decreases almost
linearly from the maximum to the minimum, as seen from the dash-dotted line in Fig.~\ref{figu5}. 
Therefore, we have that
for every rotation rate there will be a stationary state corresponding to an energy saddle. We may
follow the current `trajectory' of such a saddle in Fig.~\ref{figu5} starting from $f=0$ with the $\pi$-state saddle denoted by the dashed line. Such a line continues to the maximum (circle), where a bifurcation takes place
and the $\pi$-state (dashed line) continues
as an energy minimum, while the saddle depicted as the dash-dotted line becomes a $\phi$-state (actually, the
bifurcation yields a couple of symmetrical saddles at $\phi=\pm\arccos(K/P_{\rm eff})$ \cite{jezek21}). 
In fact, following such
a saddle we observe that a rapid decrease from the maximum is triggered, both in current and phase difference, 
with the current represented as the practically straight dash-dotted line. In such a stage, 
the phase $\phi$ decreases from
the value $\pi$ at the maximum current, passing through the value $\pi/2$ for zero current at $f=f_0/2$, and reaching
a vanishing phase difference at the current minimum, where it joins the 0-state saddle `trajectory' represented by the full line. All the currents of Fig.~\ref{figu5} were calculated from the GP simulation results, except for the extreme values
denoted by the circles, that were obtained by replacing Eq.~(\ref{sol}) in (\ref{Ivsphase}).

\subsection{Critical values of imbalance and current}\label{secrit}

The critical imbalance presents two different expressions obtained from the GBH model, 
which depend on the value of the rotation rate 
\cite{jezek21}. For frequencies outside the central interval of Fig.~\ref{figu3} the critical imbalance reads,
\begin{equation}
Z_c=\sqrt{\frac{8|K|}{NU_{\rm eff}}},
\label{zc1}
\end{equation}
whereas for frequencies within such an interval we have,
\begin{equation}
Z_c^\pm=\sqrt{\frac{-2}{P_{\rm eff} NU_{\rm eff}}(P_{\rm eff}\mp K)^2},
\label{zcpm}
\end{equation}
where $Z_c^+$ and $Z_c^-$ denote the critical imbalance of the 0- and $\pi$-modes, respectively.
Such critical values were also obtained from GP simulations as follows. For rotational frequencies
outside the central interval of Fig.~\ref{figu3}, we run real-time GP simulations starting from a TM wave function 
(\ref{psiTM}) with $Z=0$ and a phase difference $\phi$ close to the saddle value 0 or $\pm\pi$ (see, e.g., Fig.~\ref{figu10} (a)). 
On the other hand, for rotation
frequencies within the central interval, we followed the same procedure, but for
a couple of initial phase differences just at both sides of the saddle value,
$\pm\arccos(K/P_{\rm eff})$, which led to the two critical values.
\begin{figure}
\includegraphics{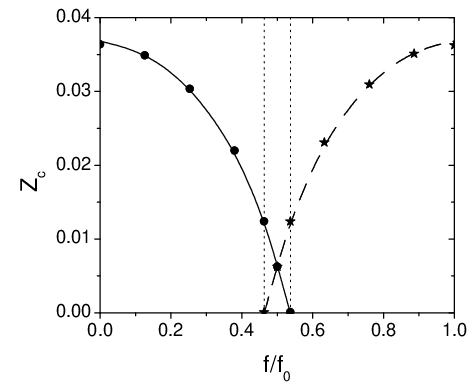}
\caption{Critical imbalance of the 0- and $\pi$-modes for the condensate of radius $r_0=3.85\,\mu$m. 
The full line (0-modes) and the dashed line ($\pi$-modes) were
obtained from  Eqs.~(\ref{zc1}) and (\ref{zcpm}), while the circles (0-modes) and stars ($\pi$-modes) correspond to the GP simulation results. The vertical dotted lines are the same as in Fig.~\ref{figu3}.}
\label{figu6}
\end{figure}
Thereby, the results from Eqs.~(\ref{zc1}) and (\ref{zcpm}) were contrasted to the GP simulation outcomes,
finding an excellent agreement for any of the condensates, as shown in Fig.~\ref{figu6}.

\begin{figure}
\includegraphics{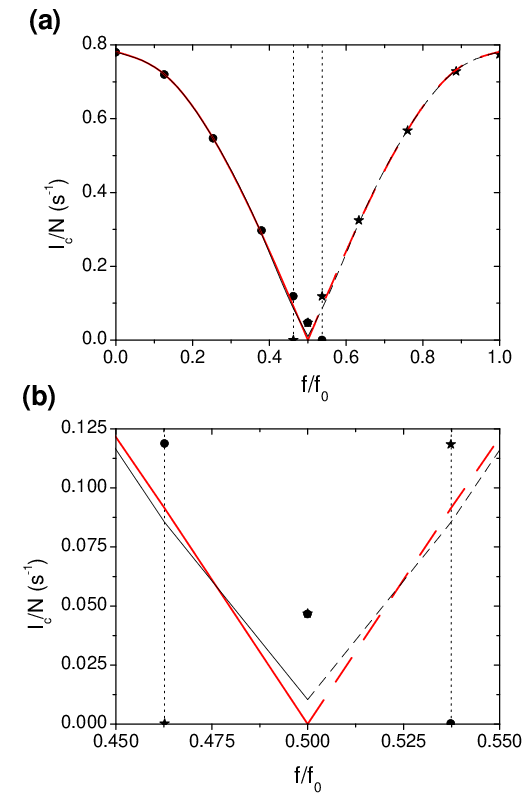}
\caption{Critical currents of the zero and $\pi$ modes for the condensate of radius $r_0=3.85\,\mu$m. 
The thin full line (0-modes) and the thin dashed line ($\pi$-modes) were
obtained by replacing Eq.~(\ref{phipmsag}) in (\ref{sagdotZ}) (only stable modes were considered),
 while the corresponding thick (red) lines arise from Eq.~(\ref{apIc}).
The circles (0-modes) and stars ($\pi$-modes) correspond to the GP simulation results, while
the vertical dotted lines are the same as in Fig.~\ref{figu3}. Graph (b) shows an enlarged view of the central bottom part of
graph (a).}
\label{figu9}
\end{figure}

\begin{figure}
\includegraphics{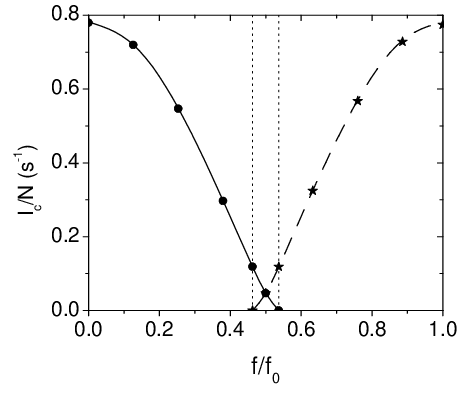}
\caption{Critical currents of the zero and $\pi$ modes for the condensate of radius $r_0=3.85\,\mu$m. 
The full-line 0-modes and the dashed-line $\pi$-modes were
obtained by replacing Eq.~(\ref{phipm}) in (\ref{zp}), while the circles (0-modes) and stars ($\pi$-modes) belong to the GP simulation results. The vertical dotted lines are the same as in Fig.~\ref{figu3}.}
\label{figu7}
\end{figure}

The critical current $I_c$ is defined as the maximum current that can flow from the upper
to the lower portion of the condensate at a given rotational frequency, i.e., the maximum of $N\dot Z/2$
with $\dot Z$ given by Eq.~(\ref{dotZ}). The dependence of the critical current on such a frequency
can act as a rotation sensor through the Sagnac effect \cite{satop,satocr,gautier}. In fact, taking into
account
the Sagnac phase shift, the phase $\xi$ in (\ref{xi}) can be approximated as,
\begin{eqnarray}
\xi & \simeq & 2n\pi -\frac{m}{\hbar}\,{\bf \Omega}\cdot {\bf\cal A}\nonumber\\
& \simeq & 2n\pi -\pi\frac{f}{f_0},
\label{sagnac}
\end{eqnarray}
where ${\bf \Omega}$, ${\bf\cal A}$ and $n$ respectively denote the angular velocity, the loop area vector and the winding number. Here we have assumed the 1D approximation $f_0\simeq\hbar/(2\pi m r_0^2)$, although
we will continue henceforth utilizing the numerical values of $f_0$ given in Table \ref{tab1}.
Now, replacing Eq.~(\ref{sagnac}) in (\ref{dotZ}) we have,
\begin{equation}
\hbar\dot Z = -4\hbar\frac{I_0}{N}[\cos(\pi f/f_0) \sin\phi
+\alpha_0\cos(2\pi f/f_0) \sin 2\phi ].
\label{sagdotZ}
\end{equation}
It is easy to find that the extremes of Eq.~(\ref{sagdotZ}) correspond to the following phase differences,
\begin{eqnarray}
\phi&=&\arccos\left\{-\frac{\cos(\pi f/f_0)}{8\,\alpha_0\cos(2\pi f/f_0)}\right.\nonumber\\
&\pm&\left.\sqrt{\frac{1}{2}+\left[\frac{\cos(\pi f/f_0)}{8\,\alpha_0\cos(2\pi f/f_0)}\right]^2}\right\}.
\label{phipmsag}
\end{eqnarray}
The critical current obtained by replacing Eq.~(\ref{phipmsag})
in (\ref{sagdotZ}) is shown in Fig.~\ref{figu9}, along with the corresponding simulation results,
$I_c/N=\max\left\{\int dx |\Psi|^2[(\hbar/m)\partial(\arg\Psi)/\partial y-\Omega x]\right\}$,
obtained by following the same procedure employed for extracting the critical imbalances.
We also depict the approximate expression for $\alpha_0\to 0$,
\begin{equation}
\frac{I_c}{N} \to 2\frac{I_0}{N}|\cos(\pi f/f_0)|.
\label{apIc}
\end{equation}
We may see from Fig.~\ref{figu9} that both expressions yield quite accurate estimates as compared to the simulation
results, except in the neighborhood of the central frequency interval where both modes coexist. Particularly, we may observe 
in Fig.~\ref{figu9} (b) that the effect of neglecting the finite value of $\alpha_0$ leads to a vanishing estimate for the critical current at $f=f_0/2$, in contrast to the simulation outcome and the theoretical result with the finite $\alpha_0$.
These differences with the simulation results are probably due to the fact that a phase drop arising from the superfluid hydrodynamic inductance of the ring has been neglected \cite{nara}. Such a phase would represent an additional term on the right-hand side of Eq.~(\ref{sagnac}) given by $-(\pi\beta_L/2I_0)(I_l+I_r)/2$, where the parameter $\beta_L$ is analogous to the screening parameter in SQUIDs \cite{gross2016}.
Nonetheless, we will see that the above discrepancies between simulation and theoretical estimates can be solved within the
GBH model. 
In fact, taking into account that $Z^2\ll 1$ for any frequency (see Fig.~\ref{figu6}), we obtain that the extremes of the right-hand side of Eq.~(\ref{zp}) may occur for the following two values of the phase difference,
\begin{equation}
\phi=\arccos\left[\frac{K}{4\,P_{\rm eff}}\pm\sqrt{\frac{1}{2}+\left(\frac{K}
{4\,P_{\rm eff}}\right)^2}\right],
\label{phipm}
\end{equation}
where it is easy to show that only one of the above values yields a valid result, except for rotation frequencies
within the central interval of Fig.~\ref{figu3}, where both results actually yield valid phase differences.
We depict in Fig.~\ref{figu7} the values of the critical current obtained by replacing Eq.~(\ref{phipm}) in (\ref{zp}),
where we may observe that, similarly to Fig.~\ref{figu6}, there is an excellent agreement between the GBH model
and the simulation results for the whole frequency range. Such a good behavior is reproduced as well for the remaining condensates.
It is also interesting to note from Fig.~\ref{figu7} that the rotational sensitivity $\Delta I_c/\Delta f$ of the AQUID
grows with $f_0^{-1}\sim r_0^2$, i.e., the area enclosed by the ring \cite{satop,satocr,gautier}.

\begin{figure}
\includegraphics{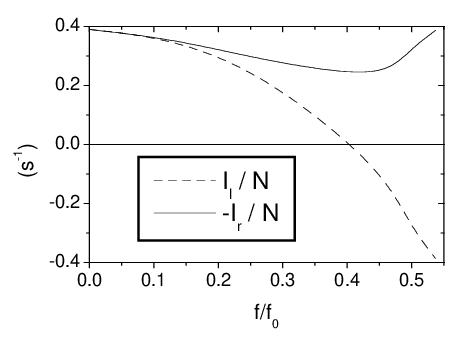}
\caption{Components from the left and right JJ of the critical current of the 0-mode in
the condensate of radius $r_0=3.85\,\mu$m in Fig.~\ref{figu1} (a).
The sum of such currents flowing from above to below across the left JJ ($I_l$, dashed line) and 
the right JJ (-$I_r$, full line) yields the
corresponding critical current of Fig.~\ref{figu7}.}
\label{figu8}
\end{figure}

The dependence of the critical current on the rotation frequency seen in Fig.~\ref{figu7},
arises as a result of the interference of the currents flowing across each JJ.
Specifically, the current oscillations from each JJ having the same amplitude given by the fifth column of Table \ref{tab3}, are generally out of phase, with a phase difference that depends on
the rotation rate as the Sagnac phase shift (\ref{sagnac}), approximately. 
This is actually the phenomenon behind the name of a `quantum interference device' in the AQUID and,
more generally, in all SQUIDs. In Fig.~\ref{figu8}, we depict both components
of the critical current of the 0-mode in Fig.~\ref{figu7}
stemming from the left and right JJ.
Note that such contributions range from being exactly the same (constructive
interference) for the non-rotating condensate, to becoming just opposite
to each other (destructive interference)
at the frequency where the 0-mode disappears by transitioning from an energy minimum to a saddle.

Finally, we display in Table \ref{tab4} the maximum and minimum values of the critical imbalance and current for each condensate. Note that the maximum currents of the fifth column almost coincide with twice the values of the same column in Table \ref{tab3}, as expected. As regards the minimum values displayed in Table \ref{tab4},
they correspond to the stable states at $f=f_0/2$, where both lines intersect in Figs.~\ref{figu6} and \ref{figu7}.
\begin{table*}
\caption{ Maximum and minimum values of the critical imbalance and current for each condensate.}
\begin{ruledtabular}
\begin{tabular}{lccccc}
$r_0$ ($\mu$m)  &   $\mu_{\rm GS}/V_b$ &   $\max(Z_c)$ &  $\min(Z_c)$ &  $\max(I_c/N)$ (s$^{-1}$) &  $\min(I_c/N)$ (s$^{-1}$) \\
\hline
 3.85 &  0.876 &  0.03637 &   0.00626  & 0.7799 & 0.04657 \\

 4.82 &  0.717 &  0.01739 &   0.00155  & 0.1406 & 0.00222  \\

 8.00 &  0.677 &  0.01071 &   0.00103  & 0.0507 & 0.00096  \\
 
 8.00 &  0.640 &  0.00924 &   0.00074  & 0.0347 & 0.00044  \\
\end{tabular}
\end{ruledtabular}
\label{tab4}
\end{table*}

\subsection{Quantum superposition of persistent-current states and atom number parity-protected qubit}\label{sec6}
The semiclassical Hamiltonian from which the modified TM equations (\ref{TMeqsm}) arise via Eqs.~(\ref{form})
reads,
\begin{equation}
\hbar{\cal H}= U_{\rm eff}{\cal N}^2
- NK\cos \phi+ N\frac{P_{\rm eff}}{4}\cos 2\phi,
\label{hamclass}
\end{equation}
where we have assumed $({\cal N}/N)^2\ll 1$ (see, e.g., Fig.~\ref{figu6}). 
The stable stationary states treated so far correspond to minima of the condensate energy represented by the above
Hamiltonian. However, such a treatment is basically classical, since it overlooks any quantum feature of the stationary states. Thereby, to reveal such characteristics one should proceed to quantize the expression (\ref{hamclass})
\cite{stringari}.
This arises immediately by replacing the conjugate variables $\phi$ and ${\cal N}$ with operators 
satisfying the commutation relation $[\hat{\phi},\hat{\cal N}]=i$. 
Thus, in the phase representation where
$\hat{\cal N}=-i\partial /\partial\phi$, we obtain the following Hamiltonian acting in the space of periodical functions
of period $2\pi$,
\begin{equation}
\hat{\cal H}= -U_{\rm eff}\frac{\partial^2}{\partial\phi^2}+V_f(\phi),
\label{Hper}
\end{equation}
where the potential energy
\begin{equation}
V_f(\phi)=- NK\cos \phi+ N\frac{P_{\rm eff}}{4}\cos 2\phi
\label{Vf}
\end{equation}
turns out to be dependent of the rotational frequency $f$ through the GBH 
parameters $K$  and $P_{\rm eff}$ (recall that we have neglected such a dependence in the on-site interaction
parameter $U_{\rm eff}$). 
\begin{figure}
\includegraphics{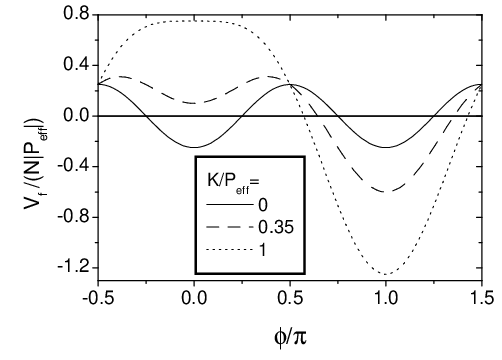}
\caption{Potential energy (\ref{Vf}) for three values of the quotient $K/P_{\rm eff}$.}
\label{figu11}
\end{figure}
We display in Fig.~\ref{figu11} the above potential energy for three values of
the quotient $K/P_{\rm eff}$. The null value of this quotient corresponding to the central frequency $f=f_0/2$
in Fig.~\ref{figu3}, yields the full-line sinusoid of
a symmetric doble-well potential with minima
at $\phi=0$ and $\pi$ in Fig.~\ref{figu11}. 
Moving away from such a central frequency, values within the central interval between the vertical dotted lines in
Fig.~\ref{figu3}, yield tilted double-well potentials, as that denoted by the
dashed line in Fig.~\ref{figu11} for a frequency above the central value $f_0/2$. Such a tilted-double-well-potential pattern
ends at the borders of the central frequency interval of Fig.~\ref{figu3}, where $|K/P_{\rm eff}|=1$ and
one of the energy minima becomes a saddle, 
as observed in Fig.~\ref{figu11} for the dotted-line potential of the rotational
frequency $f=(f_0+\Delta f)/2$.

\begin{figure*}
\includegraphics{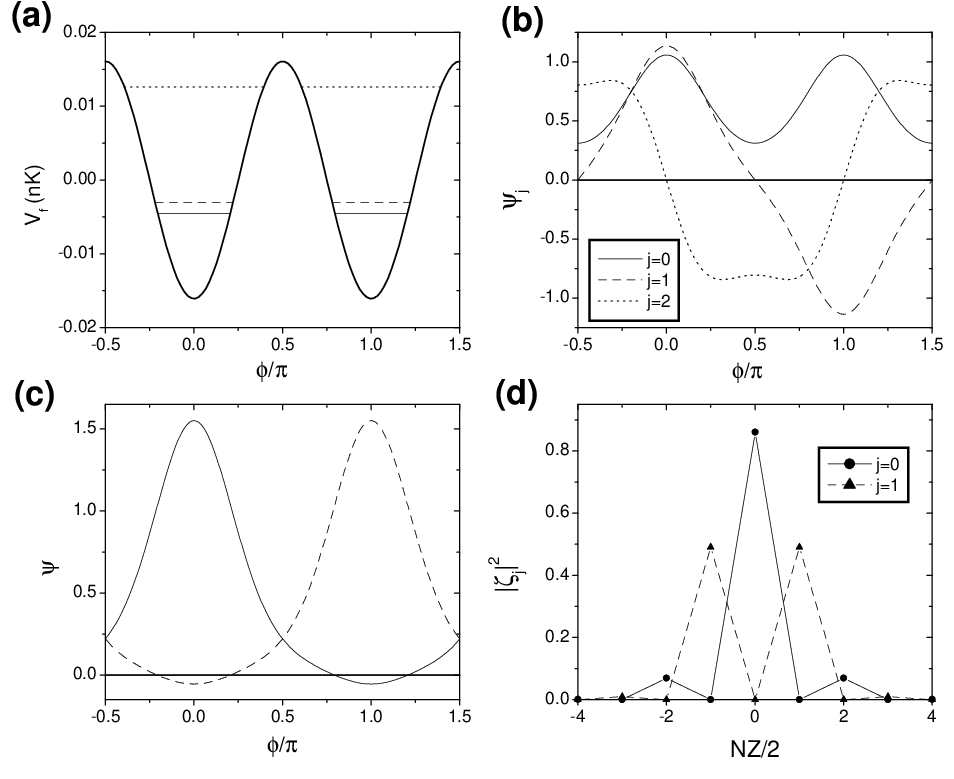}
\caption{Lowest eigenstates of Hamiltonian (\ref{Hper}) for the
condensate with $r_0=8\,\mu$m and $\mu_{\rm GS}/V_b=0.677$
at the rotational frequency $f=f_0/2$.
Symmetric double-well potential $V_f$ (thick full curve) and the first three energy levels (a),
with their respective eigenfunctions (b). (c) Localized `persistent-current'
wave functions $\psi_-$ (full line) and $\psi_+$ (dashed line).
(d) Squared norm of the wave function (\ref{zeta}) for the qubit states. We have assumed an even
number of particles $N$, so only integer numbers 	${\cal N}=NZ/2$ form the domain of the wave functions
$\zeta_j(\cdot)$, and the lines between symbols were represented only to guide the eye.}
\label{figu13}
\end{figure*}

Now we begin by focusing on the eigenvalue problem of Hamiltonian (\ref{Hper})
for the simplest case of a vanishing single-particle tunneling parameter $K$, which occurs for the
rotational frequency $f=f_0/2$. As seen in Appendix \ref{appB}, the eigenfunctions of such a Hamiltonian
correspond to the set of Mathieu functions of the first kind. Figure~\ref{figu13} (a) displays the first
three energy levels of the condensate with radius $r_0=8\,\mu$m and $\mu_{\rm GS}/V_b=0.677$, while panel
(b) shows the corresponding eigenfunctions $\psi_j(\phi)$, namely $\psi_0(\phi)={\rm ce}_0(\phi +\pi/2,q)$, 
$\psi_1(\phi)={\rm se}_1(\phi +\pi/2,q)$, and $\psi_2(\phi)={\rm ce}_1(\phi +\pi/2,q)$, with $q=1.300$, written
in terms of cosine- and sine-elliptic functions.
Note that these first eigenfunctions exhibit 
the expected form consistent with the symmetric double-well potential of panel (a). We will study
the eigenstates of the first two levels as possible candidates for a qubit. To minimize decoherence,
it is a common requirement to have an energy spectrum with a gap from the qubit levels to the third eigenvalue. To quantitatively evaluate such a characteristic, we introduce
a qubit quality factor \cite{3-wl}, $Q=\Delta E_2/\Delta E_1$ (with $\Delta E_j$ the energy of the $j$-th eigenstate
relative to the ground state), which should be greater than that of three equidistant levels, $Q=2$. 
We may see from Table \ref{tab5} that this condition is fulfilled by all the condensates considered in our study,
particularly, that of Fig.~\ref{figu13} (a), which has $Q=11.10$. 
\begin{table*}
\caption{Qubit parameters of the different condensates. The values in parentheses for ${\cal T}$ stem
 from the approximation (\ref{deltaa}).}
\begin{ruledtabular}
\begin{tabular}{lccccccc}
$r_0$ ($\mu$m)   &   $\mu_{\rm GS}/V_b$  & $q$ & $Q$ & ${\cal T}$ (s) & $(\Delta f)_{\rm eqd}/f_0$ &
$(\Delta f)_{\rm p-p}/f_0$ & $(\Delta f)_{\rm p-p}/f_0\times Q$ \\
\hline\\
 3.85  &  0.876 &  22.839 & $1.425\times 10^{7}$  & $1.61\,(1.53)\times 10^{6}$  & 0.002 & 2.2$\times$10$^{-10}$ & 0.00314\\
 4.82  &  0.717 &  1.100 & 7.968  & $7.34\, (5.50)$ & 0.002 & 0.00032 & 0.00255\\
 8.00  &  0.677 &  1.300 & 11.10  & $15.65\, (12.08)$ & 0.0022 & 0.00024 & 0.00266\\
 8.00  &  0.640 &  0.5086 & 2.726  & $6.44\,(4.26)$ & 0.0015 & 0.00107 & 0.00292\\
\end{tabular}
\end{ruledtabular}
\label{tab5}
\end{table*}
Table \ref{tab5} also shows that the quality factor
$Q$ seems to be an increasing function of the $q=N|P_{\rm eff}|/(8U_{\rm eff})$ parameter of the Mathieu equation.
However, later we will see that it is not the whole story to achieve the qubit with the highest possible value of the
$q$ parameter, or the quality factor, since such a setup might require keeping the rotation frequency at the value $f=f_0/2$ within an unattainable experimental precision.

In Fig.~\ref{figu13} (c) we depict the wave functions $\psi_\mp=(\psi_0\pm\psi_1)/\sqrt{2}$ of localized states around
the phases 0 and $\pi$ corresponding to both `classical' persistent-current states of Fig.~\ref{figu5} (the subscript of $\psi_\mp$ indicates the sign of the corresponding current).
However, such states are not quantum mechanically stationary since they are built from a superposition of two eigenstates
of different energy. An initial state of this kind would evolve in time oscillating between both `persistent-current' states at the period ${\cal T}=\hbar\pi/\Delta E_1$. Note that only for the highest value of
the chemical potential in Table \ref{tab5}, the nearly degenerate energy
levels of the qubit yield truly persistent currents with a huge value of the period ${\cal T}$.
Actually, according to Eq.~(\ref{deltaa}), the gap between both qubit levels
decreases exponentially with increasing $\sqrt{q}=\sqrt{N|P_{\rm eff}|/(8U_{\rm eff})}$.
Even a faster decrease with the number of particles has been predicted for 1D ring-shaped condensates with a single
barrier for intermediate \cite{hw} and strong \cite{anun} interactions, a fact which would severely limit the size of superposition states in any realizable experimental system. In our case, however,  the dependence on $\sqrt{N}$ of $\Delta E_1$ yields a more favorable scaling, where mesoscopic superpositions involving thousand of atoms are expected
to be feasible. Note that the qubit states arise as symmetric and anti-symmetric superpositions of both states $\psi_+$ and $\psi_-$ \cite{solenov,hw,anun,anun0},
\begin{subequations}
\begin{equation}
\psi_0=(\psi_-+\psi_+)/\sqrt{2}
\end{equation}
\begin{equation}
\psi_1=(\psi_--\psi_+)/\sqrt{2}.
\end{equation}
\label{eig}
\end{subequations}

Valuable
complementary information regarding the eigenstates can be obtained from the wave functions in the `momentum' 
${\cal N}=NZ/2$ representation, $\zeta({\cal N})$, that arise as usual from the Fourier transform of the `coordinate' 
wave functions $\psi(\phi)$,
\begin{equation}
\zeta({\cal N})=\frac{1}{\sqrt{2\pi}}\int_{-\frac{\pi}{2}}^{\frac{3}{2}\pi}\exp(-i\phi{\cal N})\,\psi(\phi)\,d\phi.
\label{zeta}
\end{equation}
Figure~\ref{figu13} (d) shows the probability $|\zeta({\cal N})|^2$ of having tunneled ${\cal N}$ bosons
from the upper to the lower portion of the condensate for both states of the qubit. For the ground-state,
we observe that the maximum probability corresponds to ${\cal N}=0$, with next nonvanishing small probabilities at
${\cal N}=\pm 2$. Such a distribution can be understood as describing the tunneling fluctuations of a ground-state
with $N/2$ bosons at each portion of the condensate, and, since only tunneling of pairs are allowed at 
the frequency $f=f_0/2$,
all odd values of ${\cal N}$ will have probability zero, as observed in Fig.~\ref{figu13} (d). On the other hand,
the first-excited eigenstate shows two symmetrical maxima at ${\cal N}=\pm 1$ and vanishing probabilities
for even ${\cal N}$. Thus, similarly to the ground-state, we may interpret this result by concluding that the first-excited wave function describes
the tunneling fluctuations of a configuration with $N/2\pm 1$ particles at each portion of the
condensate. Therefore, we have that both states of the qubit will have a different parity of the number of bosons
at each portion, a parity that will be conserved against fluctuations. In other words, similarly to
Majorana qubits and $\pi$-periodic JJs in superconducting circuits \cite{larsen20,smith20,glad,doucot}, 
the quantum information of the qubit will be protected by using such disconnected parity subspaces.

\begin{figure*}
\includegraphics{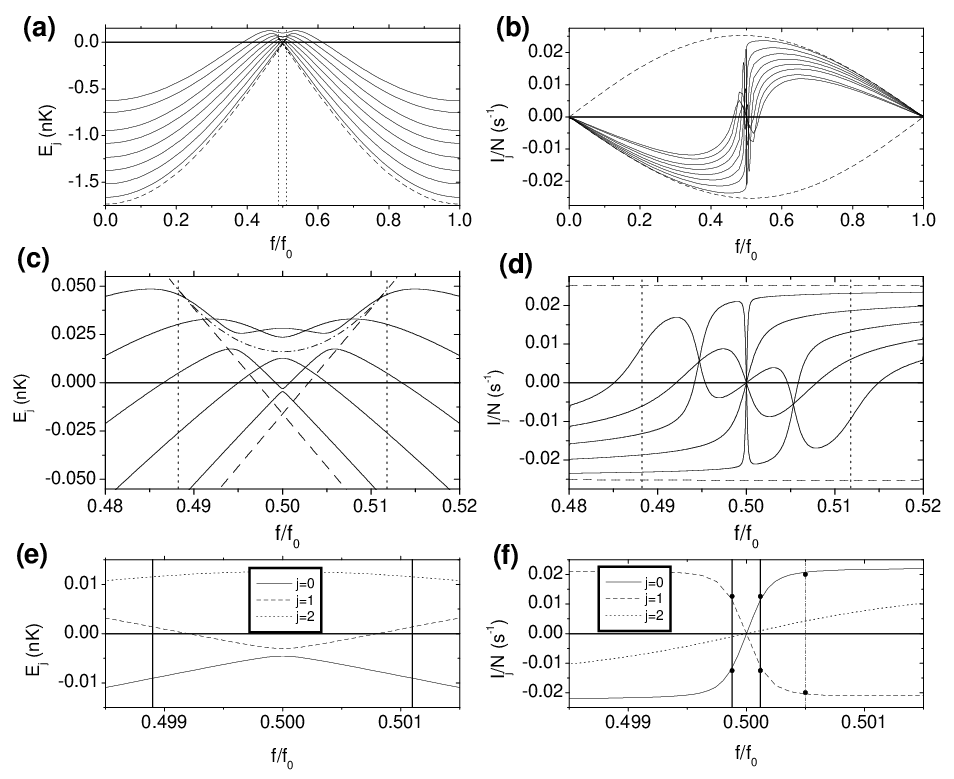}
\caption{Energy levels and stationary currents of the
condensate with $r_0=8\,\mu$m and $\mu_{\rm GS}/V_b=0.677$.
(a) First eight energy levels (full lines) and minima of the potential energy $V_f$ (dashed lines).
The left (right) dashed line corresponds to the minimum at $\phi=0$ ($\phi=\pi$). The vertical dotted lines delimit
the frequency range where both minima coexist. (b) Stationary currents corresponding to the energy levels of (a)
(full lines) and `classical' stationary currents of the zero and $\pi$ states (dashed lines) (cf. Fig.~\ref{figu5}).
(c) Enlarged view of the top central part of graph (a) showing the first five energy levels (full lines),
and the minima (dashed lines) and maximum (dash-dotted line) of the potential energy $V_f$ (cf. Fig.~\ref{figu11}).
(d) Enlarged view of the central part of graph (b) showing the stationary currents corresponding to the energy levels of (c).
The vertical dotted lines
are the same as those in graphs (c) and (a). Panels (e) and (f) correspond to enlarged views of the central parts of graphs (c) and (d), respectively, showing the first three energy levels and their respective currents. The vertical thick full lines delimit the frequency intervals
$(\Delta f)_{\rm eqd}$ in (e) and $(\Delta f)_{\rm p-p}$ in (f) (see text for explanation). The vertical dash-dotted
line in (f) indicates the rotational frequency $0.5005\,f_0$, whereas the circles correspond to values arising from Eqs.~(\ref{mcurr}).}
\label{figu12}
\end{figure*}

According to what explained in Appendix \ref{appB}, we have numerically solved the eigenvalue problem
of Hamiltonian (\ref{Hper}) for the whole frequency range. 
The energy levels for a condensate with $r_0=8\,\mu$m 
and $\mu_{\rm GS}/V_b=0.677$ are shown in Fig.~\ref{figu12} (a). 
From these levels $E_j$, one can derive the corresponding stationary currents $I_j$ as \cite{goldman,bloch},
\begin{equation}
I_j=-\frac{f_0}{2\pi\hbar}\frac{\partial E_j}{\partial f},
\label{Ibloch}
\end{equation}
where we depict them in Fig.~\ref{figu12} (b). The frequency range where the potential energy $V_f$ presents
 two energy minima, which is delimited by the vertical dotted lines in Fig.~\ref{figu12} (a), 
 deserves an enlarged view shown in panels (c) and (d).
In fact, we note in (c) a number of level crossings for the first five energy levels, of particular interest being the crossings of the second and third level, which naturally set
an upper limit for the amplitude of the frequency interval that can host a central qubit. 
Such an estimate, however, should be reduced if we wish to preserve the value of the qubit quality factor $Q$ above two. We show 
in panel (e) such a reduced
interval $(\Delta f)_{\rm eqd}/f_0$, which is located between the vertical full lines corresponding to
the frequencies where the first three energy levels turn out to be equidistant. We may compare in Table \ref{tab5} the
length of this interval for the different condensates, where we observe only small variations.

Panel (d) shows a clear qualitative difference for the behavior of the currents within
and outside the frequency interval where both energy minima coexist. That is, we observe 
that within such an interval, the 
currents appear quite entangled around the vanishing value of the central crossing point at $f=f_0/2$, 
whereas outside this interval they become completely disentangled, tending to
recover the
`semiclassical' aspect of a parallel bunch of curves that follow the persistent-current patterns of the zero and $\pi$ states
shown in panel (b).

\begin{figure}
\includegraphics{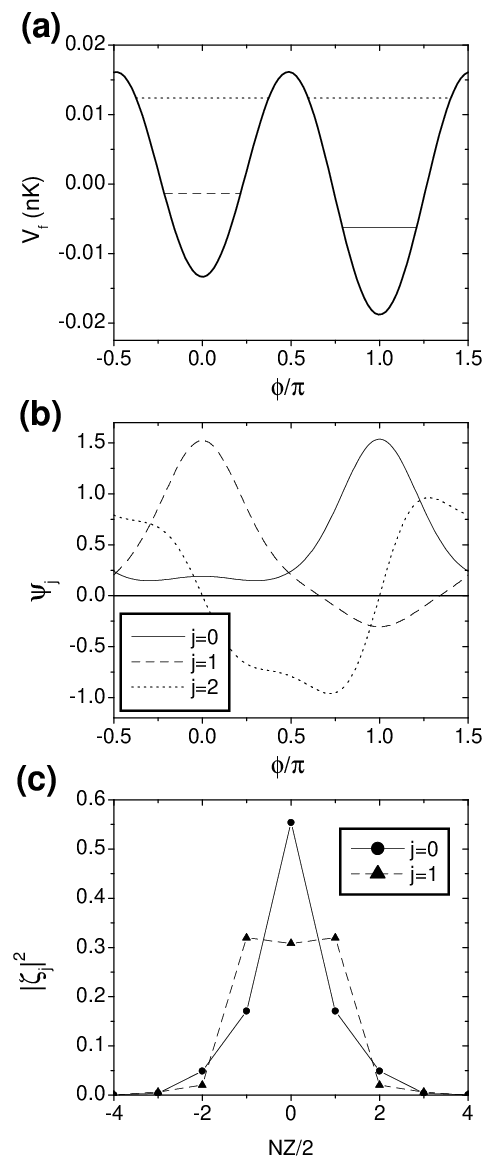}
\caption{Lowest eigenstates of Hamiltonian (\ref{Hper}) for the
condensate with $r_0=8\,\mu$m and $\mu_{\rm GS}/V_b=0.677$
at the rotational frequency $f=0.5005\,f_0$.
Tilted double-well potential $V_f$ (thick full curve) and the first three energy levels (a),
with their respective eigenfunctions (b).
(c) Squared norm of the wave function (\ref{zeta}) for the first two eigenstates. 
Similarly to Fig.~\ref{figu13} (d) the straight lines between symbols were depicted only to guide the eye.}
\label{figu14}
\end{figure}

\begin{figure}
\includegraphics{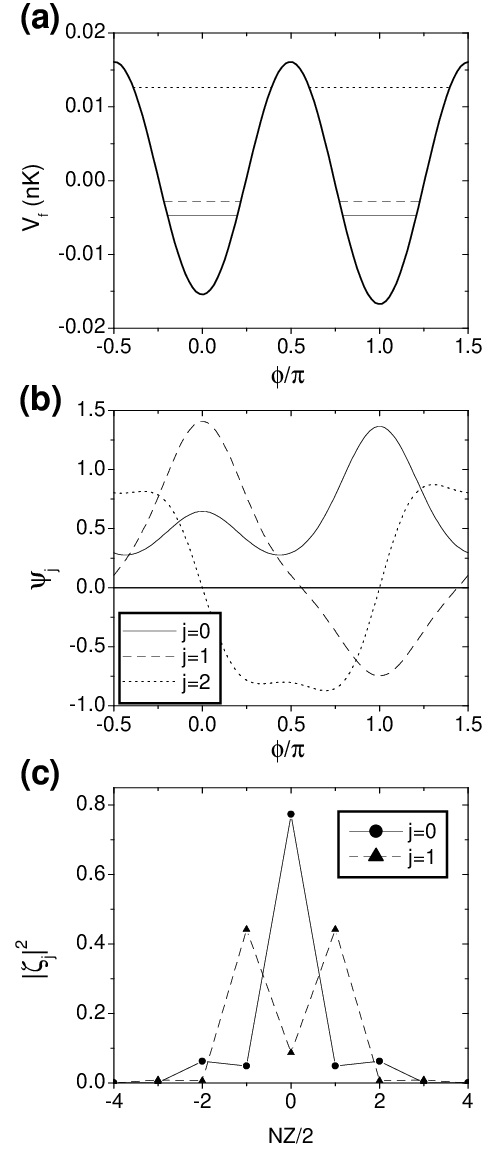}
\caption{Same as Fig.~\ref{figu14} for a rotational frequency $f=0.50012\,f_0$.}
\label{figu15}
\end{figure}

Next we will try to estimate the error within which the rotational frequency $f_0/2$ of the qubit should be determined
in order to preserve an acceptable level of parity protection. In Fig.~\ref{figu14}, we have represented the three
lowest eigenstates of the condensate of Fig.~\ref{figu13}, but at the
slightly higher rotational frequency $f=0.5005\,f_0$. 
We note that in contrast to Fig.~\ref{figu13}, now the qubit eigenfunctions become quite well localized around
each minimum of the potential energy, as depicted in the level plot of Fig.~\ref{figu14} (a), drawn
according to what observed in panel (b). Such a localization corresponds to persistent-current qubit states with opposite polarity, as observed at the vertical dash-dotted line in Fig.~\ref{figu12} (f).
The two circles on such a line represent the values arising from Eqs.~(\ref{mcurr}), which show a good agreement 
with those obtained from Eq.~(\ref{Ibloch}) (full and dashed curves).
On the other hand, Fig.~\ref{figu14} (c)
shows that the parity-protected scheme of Fig.~\ref{figu13} (d) has now been completely removed. 
It is clear then, that in order to achieve an acceptable degree of parity protection, we should consider 
values of the rotational frequency
closer to $f_0/2$. This is the case of Fig.~\ref{figu15}, where we have reduced such a frequency to the value
$0.50012\,f_0$. 
In fact, we may see in panel (c) that the parity-protected scheme appears now only slightly affected, 
as compared to Fig.~\ref{figu13} (d). On the other hand, panels (a) and (b) show that the eigenfunctions are
appreciably extended around both potential energy minima, with stationary currents built from
a quantum superposition of both persistent-current states as predicted by Eqs.~(\ref{mcurr}) (see the circles on
the right vertical full line at $f=0.50012\,f_0$ in Fig.~\ref{figu12} (f)). This result suggests the definition of a `parity-protected' frequency interval around the qubit value $f=f_0/2$, within which the `tilted qubit' states are built from at least a minimum of quantum superposition of both persistent-current states. To be precise, within such an
interval both components in Eqs.~(\ref{eigen}) should have values above a certain minimum in order to preserve
the parity-protected scheme. In fact, within the frequency interval 
delimited by the vertical full lines in Fig.~\ref{figu12} (f), the probabilities $A^2$ and $1-A^2$ 
of both persistent-current states in Eqs.~(\ref{eigen}) have values above 0.2, which correspond to
mean current values of magnitude less than 0.6$I_p$ in Eqs.~(\ref{mcurr}) (see the circles on both vertical full lines).
Thus, we may see that the above parity-protected frequency interval $\Delta f_{\rm p-p}$ can be determined by defining  the fraction of the plateau current $I_p$ below which the magnitude of mean currents of tilted qubit states should stay.
We will adopt the above fraction of 0.6 throughout this paper, since we will see that
no qualitative aspect of our discussion will depend on the precise value of such a fraction. Table \ref{tab5} displays
the value of the normalized interval
$\Delta f_{\rm p-p}/f_0$ for the different condensates studied in this work. Here it is important to
remark that all condensates have shown similar levels of parity protection within the frequency interval 
$\Delta f_{\rm p-p}$. We observe a very interesting feature
from the last column of Table \ref{tab5}, where we display the product of the quality factor $Q$ and the
normalized value of the parity-protected interval $\Delta f_{\rm p-p}/f_0$, which, despite the quite different values
of both parameters, with orders of magnitude of difference for the different condensates, remains rather constant at a
value $\sim$0.003, i.e.,
\begin{equation}
\Delta f_{\rm p-p}/f_0\simeq\frac{0.003}{Q}.
\end{equation}
The above equation shows that the quality factor of the qubit and the error interval within which the rotation frequency $f_0/2$ must be experimentally determined, turn out to be closely related.
That is, a high value of the quality factor might be eventually unreachable, since it could require to maintain the frequency value $f_0/2$ within an unattainable experimental precision. This is clearly the case for the condensate with the highest
value of the chemical potential in the first row of Table \ref{tab5}. So, the way to achieve a suitable balance between the quality factor
and the maximum allowed frequency error seems to be to reduce the chemical potential value (e.g., by reducing the number of particles), as seen for the other condensates in Table \ref{tab5}. The best trade-off between both magnitudes 
seems to be given by the
qubits of the second and third row of Table \ref{tab5}
that have quality factors of order 10 with a maximum permitted frequency error of about $0.05\%$.

Finally, we note that the generation of a macroscopic quantum superposition has proven to be an extremely challenging task. Here we mention some experimental proposals to prepare and detect a superposition
of persistent-current states. In Ref.~\cite{anun0} it has been shown that such states can be distinguished in time-of-flight
absorption images and it was proposed to probe the cat-like correlations via the many-body oscillations induced by a sudden change in the qubit rotation frequency. Much more recently \cite{haug18}, 
it was discussed how a self-heterodyne protocol can be utilized to
detect states with macroscopic quantum coherence made of the above superpositions in a rf-AQUID.
Along the same lines, Ref. \cite{haug21} has shown how to engineer bosonic entangled currents of multiple winding numbers in a robust manner by using deep reinforcement learning. 

\section{Summary and Conclusion}\label{sec5}
We have developed a GBH model in the TM approximation for a dc AQUID similar to those recently studied in
Ref.~\cite{ryu20}. Taking into account the macroscopic occupation of states, the replacement of creation and
annihilation operators by complex $c$-numbers transforms the Heisenberg equations of motion into a pair of coupled
TM equations. Such equations are shown to derive from a semiclassical Hamiltonian that depends on canonically conjugate variables given by the phase difference and particle imbalance between both halves of the AQUID. The system dynamics in this model is thus ruled by the minima and saddles of a 2D energy landscape that should be in accordance with the corresponding simulation results. We have found that in order to match the energy landscape of the GBH model and that of the GP order parameter, a couple of parameters of the model must be modified. To this aim, we have derived a well-defined prescription that utilizes the period of small oscillations of GP energy minima as input.
We have shown that such a modified GBH model yields an excellent agreement to the GP simulation results for the entire rotational frequency range, reaching also the critical values of current and imbalance.
Thus, once established the above energy landscape accuracy of the modified semiclassical Hamiltonian, its quantization through standard procedures was utilized to investigate the quantum features of stable stationary states.
We have studied in this respect the whole range of rotational frequencies, which corresponds to the period $f_0$ of
stationary currents as functions of such a frequency. However, the most interesting part of the quantum behavior occurs
within the central interval of frequencies around $f_0/2$,
where the potential energy of the Hamiltonian presents two minima.
Just at the center of this interval, which corresponds to a half of magnetic flux quantum in the 
electromagnetic analogy,
such minima are symmetric and the two lowest eigenstates become
potential candidates to form a qubit. In fact, we have found that the most probable number of bosons at both halves of
the AQUID turns out to be $N/2$ for the ground state and $N/2\pm 1$ for the first-excited state. That is, the
parity of the number of particles at each portion turns out to be different for both qubit states, and moreover
such parities are robust against fluctuations since only tunneling of boson pairs is permitted for $f=f_0/2$. 
In other words, we have a parity-protected qubit similar to those of
superconducting circuits threaded by a half-quantum of applied flux \cite{larsen20,smith20,glad,doucot}.
We have explored an important aspect of the feasibility of such a qubit, which is the minimum experimental
precision that would be required to establish the value $f_0/2$ of the rotation frequency without losing the main features of the parity protection scheme. We have shown that such a scheme is preserved within a frequency interval around
$f_0/2$, where the qubit eigenstates arise from at least a minimum of quantum superposition of both persistent-current states of opposite polarity. Such an interval, which determines the maximum admissible error for the rotation frequency $f_0/2$, turns out to be very simply related to the quality factor that quantifies the gap between the energy levels of the qubit and the following levels.
In fact, being the product of both nearly a constant, a good qubit design would require achieving the best trade-off between
such qubit characteristics. Here it is worthwhile noticing that an adequate control parameter for tuning between both magnitudes is given by the chemical potential, or more simply, by the condensate number of particles.
For instance, among the configurations we have considered in our study, we have found that a quality factor of order 10 within a parity-protected regime can be achieved under an error of 0.05\% in the value of the frequency $f_0/2$, a goal that seems fully experimentally accessible. In conclusion, we have shown that a dc AQUID can work as a parity-protected qubit and, having explored important aspects of its feasibility, we hope that this study can contribute to its eventual implementation.

\acknowledgments
This work was supported by CONICET and Universidad de Buenos Aires through Grants
 PIP 11220210100821CO and UBACyT 20020190100214BA, respectively.

\appendix
\section{Modified values of GBH parameters $U$ and $P$}\label{appA}
In this Appendix we will discuss the method for obtaining the modified values of the parameters $U$ and $P$ that
is based on numerically extracting the period of small oscillations around the stable stationary states.
We note that such a procedure could be applied to any geometry of the condensate other than the present case of a ring.
However, the difference between the original and modified values will of course depend on each case.

 To extract the effective value $U_{\rm eff}$ of the on-site interaction parameter,
we have made use of the formula for the period of small
oscillations of the Hamiltonian  ${\cal H}$ in Eq.~(\ref{form}), which depends on the momentum
${\cal N}=NZ/2$ and the coordinate $\phi$ as the harmonic oscillator Hamiltonian
(\ref{so0}) with the time period,
\begin{equation}
T_0=\frac{\pi\hbar}{\sqrt{\frac{NU}{2}(K-P)}}.
\label{T0}
\end{equation}
We have numerically obtained the value of $T_0$ by running a real-time GP simulation for a non-rotating condensate, starting from an initial wave
function (\ref{psiTM}) with $\phi =0$ and $Z\ll 1$. On the other hand, 
the values of $K$ and $P$ in Eq.~(\ref{T0})
were obtained as in Fig.~\ref{figu2}. So, replacing such values in  (\ref{T0}) we were able to extract
a value of $U$, which actually corresponds to $U_{\rm eff}$.
Given the rather weak dependence of the on-site interaction $U$ on the rotation rate,
we have assumed that the above value for $U_{\rm eff}$ remains valid for rotating condensates.
We note that, unlike the previous method to obtain the effective value that was based on stationary order 
parameters \cite{cat20,cap13,nigro17,jezek21}, the current procedure, which resorts to time-dependent simulations,
should always be reliable.

To extract $P_{\rm eff}$, we have utilized the expression of the
small-oscillation period corresponding
to the Hamiltonians (\ref{so0}) and (\ref{sopi}) with the replacements 
$U\to U_{\rm eff}$ and $P\to P_{\rm eff}$. Thus, it is easy to derive from such expressions
the modified pair-tunneling parameter $P_{\rm eff}$ as,
\begin{equation}
P_{\rm eff} = \pm K-\frac{2\pi^2\hbar^2}{NU_{\rm eff}T_\pm^2},
\label{epeff}
\end{equation}
where $T_+$ and $T_-$ respectively denote the small-oscillation periods for the 0- and 
$\pi$-modes. Again, such periods are obtained by running real-time GP simulations
starting from the corresponding TM wave function (\ref{psiTM}) with $Z\ll 1$. 
The above procedure for obtaining $P_{\rm eff}$ was previously utilized in \cite{jezek21} as an alternative and
apparently
more accurate method of calculating $P$. However, it is important to remark that both calculations actually yield
 two different parameters.
 
Given that $U_{\rm eff}
> 0$, we may see from Eq. (\ref{epeff}) that the condition for the existence of an energy minimum in
a 0- or $\pi$-state should be $K>P_{\rm eff}$ or
$K<-P_{\rm eff}$, respectively. Otherwise, if such conditions are not met, the stationary
state should correspond to an energy saddle. 

\begin{figure*}
\includegraphics[width=\linewidth]{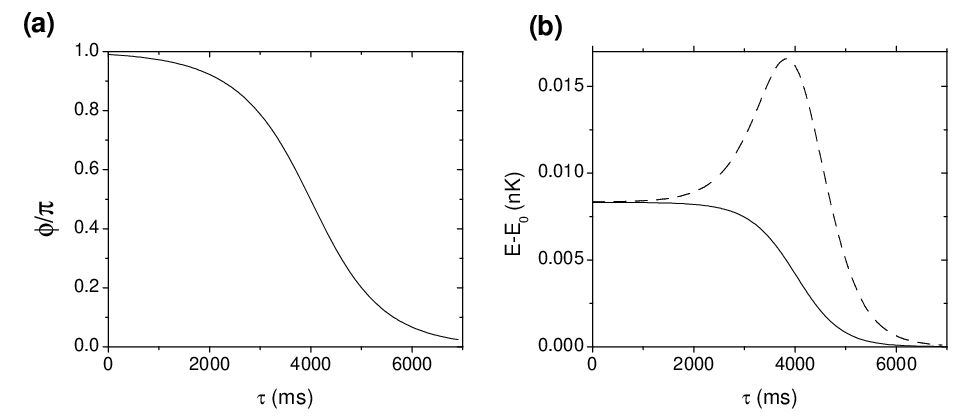}
\caption{Imaginary-time evolution of the condensate of radius
$r_0=3.85\,\mu$m for an imposed rotational frequency $f=0.253 f_0$.
The initial order parameter corresponds to a TM wave function (\ref{psiTM})
with $Z=0$ and $\phi$ close to $\pi$.
(a) Phase difference obtained from Eq.~(\ref{tau}). (b) Values of the 2D
energy functional (\ref{Efun}) above the value $E_0$ of the stationary state with $\phi=0$,
for the GP order parameter (full line) and the TM order parameter (dashed line).
The latter coincides with $E_{TM}(Z,\phi)-E_{TM}(0,0)$ from Eq.
(\ref{ener}), while the former arises from such an equation as well, but with the replacements 
$U\to U_{\rm eff}$ and $P\to P_{\rm eff}$.}
\label{figu4}
\end{figure*}

\begin{figure*}
\includegraphics[width=\linewidth]{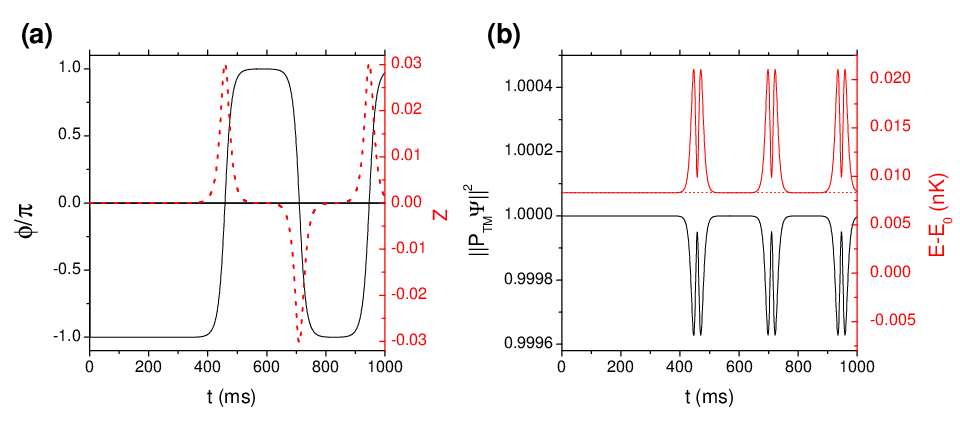}
\caption{Real-time evolution simulation of the condensate with $r_0=3.85\,\mu$m at 
an imposed rotational frequency $f=0.253 f_0$, starting
from the saddle with $\phi=-\pi$.  (a) Phase difference $\phi$ (full line) and particle imbalance $Z$ (dotted line).
(b) Squared norm of the projection of the GP order parameter onto the TM subspace $||P_{TM}\Psi||^2$
(bottom full line). Values of the
energy functional (\ref{Efun}) above the value $E_0$ of the stationary state with $\phi=0$,
for the projected order parameter $P_{TM}\Psi$ (top (red) full line) and the GP order parameter 
$\Psi$ ((red) dotted line).} 
\label{figu10}
\end{figure*}

In order to understand how the modified parameter $P_{\rm eff}$
 restore the agreement of the nature of the stationary states predicted
by the GBH model with respect to that of the simulation outcomes, we may study the behavior of the energy in the neighborhood
of such states. However, given that any real-time evolution through the GP equation should conserve the energy, it will be convenient to consider at first imaginary-time propagations. Thus, we show in Fig.~\ref{figu4} such an evolution for the condensate of radius
$r_0=3.85\,\mu$m with an imposed rotational frequency $f=0.253\, f_0$.
By comparing Figs. \ref{figu2} and \ref{figu3}, we note that while the original GBH model predicts that a 
$\pi$-state should have an energy minimum for such a frequency, 
the modified model (with $P$ replaced  by
$P_{\rm eff}$) predicts a saddle. So, we have propagated in imaginary time 
$i\tau$ an initial TM order parameter (\ref{psiTM}) with $Z=0$ and a quite small departure from $\phi=\pi$. 
Figure \ref{figu4} (a) shows the imaginary time evolution
of the phase difference $\phi$ obtained from the GP order parameter $\Psi(\tau)$ as,
\begin{equation}
\phi=\arg{\left[\frac{\int \int dx\, dy\,\,  \psi_u^*\Psi}{\int \int dx\, dy\,\,  \psi_l^*\Psi}\right]},
\label{tau}
\end{equation}
where we observe a rather slow evolution from the vicinity of the saddle point at $\phi=\pi$ to the
energy minimum at $\phi=0$. This is also seen from
the full line in Fig.~\ref{figu4} (b), which corresponds to the energy obtained from the energy functional (\ref{Efun})
with the GP order parameter. We have verified that such a curve
practically coincides with that obtained from the TM energy expression (\ref{ener}) with the replacements
$U\to U_{\rm eff}$ and $P\to P_{\rm eff}$. 
On the other hand, the values obtained from the energy
functional (\ref{Efun}) for the TM wave function (\ref{psiTM}) with $Z(\tau)$ and $\phi(\tau)$
 extracted from
the GP imaginary-time evolution, yields the dashed line of Fig.~\ref{figu4} (b), which also corresponds
to the values obtained from the TM energy (\ref{ener}). This leads us to conclude that although
the energy presents a minimum at $\phi=\pi$
for evolutions {\em within the TM subspace} spanned by the TM order parameter (\ref{psiTM}),
the actual GP evolution corresponds to a saddle that is correctly described by
the TM energy expression (\ref{ener}) with the replacements 
$U\to U_{\rm eff}$ and $P\to P_{\rm eff}$.
Moreover, we show in Sec.~\ref{secrit}
that modifying the GBH model through such replacements leads to an
excellent agreement with the GP simulation results. However, such a success could be somehow surprising, given the above discrepancy between the energy landscapes of the TM subspace and that stemming from the simulation results. Actually, such a good performance of the modified GBH model could be attributed to the fact that the GP order parameter always remains almost entirely contained within the TM subspace. To see this, we have run
a real-time GP simulation starting from the saddle, which allows to `scan' the full phase-space up to the critical imbalance
$Z_c$ ($-\pi\leq\phi\leq\pi$;
$-Z_c\leq Z\leq Z_c$), as seen from Fig.~\ref{figu10} (a). 
Thus, we have calculated the projection of the GP order parameter $\Psi$ onto the TM `plane',
\begin{eqnarray}
P_{TM}\Psi=\left[\psi_u(x,y)\int \int dx'\, dy'\,\,  \psi_u^*(x',y')\right.\nonumber\\
+\left.\psi_l(x,y)\int \int dx'\, dy'\,\,  \psi_l^*(x',y')\right]\Psi(x',y'),
\label{P_TM}
\end{eqnarray}
which according to Eq. (\ref{tau}) and
\begin{eqnarray}
Z&=&\left[\left|\int \int dx\, dy\,\,  \psi_l^*(x,y)\Psi(x,y)\right|^2\right.\nonumber\\
&-&\left.\left|\int \int dx\, dy\,\,  \psi_u^*(x,y)\Psi(x,y)\right|^2\right]\nonumber\\
&/&||P_{TM}\Psi||^2
\end{eqnarray}
with
\begin{eqnarray}
||P_{TM}\Psi||^2&=&\left|\int \int dx\, dy\,\,  \psi_l^*(x,y)\Psi(x,y)\right|^2\nonumber\\
&+&\left|\int \int dx\, dy\,\,  \psi_u^*(x,y)\Psi(x,y)\right|^2,
\end{eqnarray}
should coincide unless an overall phase and normalization factor with the TM order parameter (\ref{psiTM}).
We observe in Fig.~\ref{figu10} (b) that a small deviation from unity of the squared norm of such a projection causes a departure of the energy from the conserved GP value by more than twice such a value.
In other words, we conclude that just a tiny component of the order parameter lying outside the TM subspace can bring about appreciable differences in the energy.

\section{Diagonalization of Hamiltonian (\ref{Hper})}\label{appB}
\begin{figure*}
\includegraphics{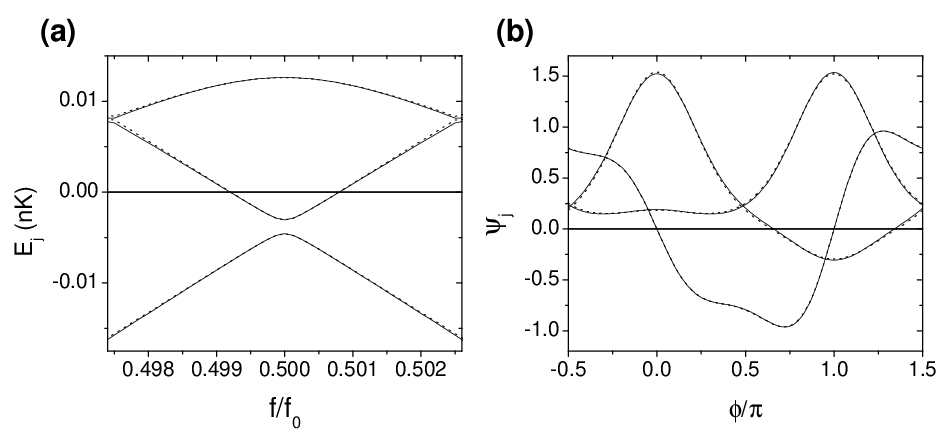}
\caption{(a) First three eigenvalues of the Hamiltonian (\ref{Hper}) for the
condensate with $r_0=8\,\mu$m and $\mu_{\rm GS}/V_b=0.677$. Panel (b) shows the respective eigenfunctions for
$f=0.5005\,f_0$ (cf. Fig.~\ref{figu14} (b)). The full lines correspond to the numerical diagonalization of a truncated
$24\times 24$ matrix, while the dotted lines represent the analytical results stemming from the diagonalization of the
blocks (\ref{block}). }
\label{figu16}
\end{figure*}

We begin by considering the simplest case of the potential energy (\ref{Vf}) for $f=f_0/2$ that has
$K=0$ and $P_{\rm eff}<0$.
Thereby, the eigenvalue equation
$\hat{\cal H}\psi(\phi) = E \psi(\phi)$ can be written as follows,
\begin{equation}
\frac{d^2f}{d\varphi^2}+[a-2q\cos (2\varphi)]f=0,
\label{mathieu}
\end{equation}
where $\varphi = \phi +\pi/2$, $f(\varphi)=\psi(\phi)$, $a=E/U_{\rm eff}$ and $q=N|P_{\rm eff}|/(8U_{\rm eff})$.
The above form (\ref{mathieu}) corresponds to 
the so-called Mathieu differential equation, which for any value of $q>0$ presents
$2\pi$-periodic solutions for a discrete spectrum of characteristic numbers $a$. Such solutions are
known as the Mathieu functions of the first kind, or more specifically, as the
 cosine-elliptic ce$_n(\varphi,q)$ and sine-elliptic se$_n(\varphi,q)$ functions of order $n$, which, sorted by
ascending characteristic numbers (or eigenvalues of $\hat{\cal H}$) are, ce$_{2k}$, se$_{2k+1}$, ce$_{2k+1}$, se$_{2k+2}$ ($k=0,1,2,...$) \cite{arscott}.
Of particular interest is the gap between the two lowest characteristic numbers $\Delta a =a_1-a_0$, which turns out
to be exponentially suppressed for large values of $q$ \cite{smith20},
\begin{equation}
\Delta a \simeq 4\sqrt{2/\pi}\, (16q)^{3/4}\exp(-4\sqrt{q}),
\label{deltaa}
\end{equation}
the above approximation being valid for $q\gtrsim 1$.
This represents
the complete solution of the eigenvalue problem of the Hamiltonian $\hat{\cal H}$ for $K=0$ and $P_{\rm eff}<0$. 
Although Mathieu's equation has often been solved numerically in the literature (see, e.g. Ref. \cite{wilk}), 
this is not the case for the Whittaker-Hill equation \cite{arscott67,*arscott70}, whose solution amounts to diagonalize
the Hamiltonian (\ref{Hper}) for finite values of both parameters $K$ and $P_{\rm eff}$ in (\ref{Vf}).
So, to find a numerical solution, we first rewrite such a Hamiltonian as $\hat{\cal H}=\hat{\cal H}_0- NK\cos \phi$, with
$\hat{\cal H}_0=-U_{\rm eff}\frac{\partial^2}{\partial\phi^2}+ N\frac{P_{\rm eff}}{4}\cos 2\phi$. 
We may then proceed to diagonalize
$\hat{\cal H}$ in the basis of eigenfunctions of $\hat{\cal H}_0$. We observe in Fig.~\ref{figu3}
that there is a wide central
frequency interval where $P_{\rm eff}<0$, in which case the eigenfunctions of $\hat{\cal H}_0$ are those that were
described above. However,
for the lowest and highest frequency intervals of Fig.~\ref{figu3}, we have instead
 positive values of $P_{\rm eff}$, for which
the $\hat{\cal H}_0$ eigenfunctions again turn out to be the Mathieu functions of the first kind,
but this time they have to be evaluated at $\varphi = \phi$. The diagonal elements of $\hat{\cal H}$ are then
the eigenvalues of $\hat{\cal H}_0$, $E^0_{j}$, while the off-diagonal matrix elements read,
\begin{equation}
E_{j,k}=-NK\int_{-\frac{\pi}{2}}^{\frac{3}{2}\pi}\psi^0_j(\phi)\,\psi^0_k(\phi)\,\cos(\phi)\,d\phi,
\end{equation}
where $\psi^0_j(\phi)$ denotes the $j$-th eigenfunction of $\hat{\cal H}_0$. We have found that
the numerical diagonalization of a truncated $24\times 24$ matrix suffices to extract the
first eight eigenvalues and eigenfunctions of $\hat{\cal H}$ within a quite good precision 
on the entire frequency interval from $0$ to $f_0$
(Fig.~\ref{figu12} (a)). There is a useful analytical approximation valid in the neighborhood of the qubit frequency $f_0/2$,
where the problem becomes reduced to a diagonalization of the following diagonal blocks of $\hat{\cal H}$,
\begin{equation}
\begin{pmatrix}
E^0_{j} & E_{j,j+1} & 0 & 0 \\
E_{j,j+1} & E^0_{j+1} & 0 & 0 \\
0 & 0 & E^0_{j+2} & E_{j+2,j+3} \\
0 & 0 & E_{j+2,j+3} & E^0_{j+3} 
\end{pmatrix},
\label{block}
\end{equation}
that yield the $\hat{\cal H}$ eigenvalues
\begin{equation}
E_{j/j+1}\simeq\frac{E^0_{j}+E^0_{j+1}}{2}\mp\sqrt{\left(\frac{E^0_{j+1}-E^0_{j}}{2}\right)^2+E_{j,j+1}^2},
\end{equation}
where the minus and plus signs in front of the square root correspond to
the energy levels $E_j$ and $E_{j+1}$, respectively. We have found that this constitutes an excellent approximation
for rotational frequencies around $f_0/2$, up to the level crossings of the second and third eigenvalue, as seen in
Fig.~\ref{figu16}.
Within the same approximation, it is easy to show that the first two eigenstates arise as quantum
superpositions of the persistent-current states $|\psi_\mp\rangle=\left(|\psi^0_0\rangle\pm
|\psi^0_1\rangle\right)/\sqrt{2}$ of Fig.~\ref{figu13} (c),
\begin{subequations}
\begin{equation}
|\psi_0\rangle=A\,|\psi_-\rangle+\sqrt{1-A^2}\,|\psi_+\rangle
\end{equation}
\begin{equation}
|\psi_1\rangle=\sqrt{1-A^2}\,|\psi_-\rangle-A\,|\psi_+\rangle
\end{equation}
\label{eigen}
\end{subequations}
with
\begin{equation}
A=\frac{E^0_1-E^0_0}{\sqrt{(E^0_1-E^0_0)^2+\left[2E_{0,1}+\sqrt{(E^0_1-E^0_0)^2+4E_{0,1}^2}\right]^2}}.
\end{equation}
The above equations (\ref{eigen}) generalize Eqs.~(\ref{eig}) for rotational frequencies in the vicinity of $f_0/2$.
Within such a frequency range, the quantum nature of the current becomes evident by introducing the observable
\begin{equation}
\hat{I}=I_p(|\psi_+\rangle\langle\psi_+|-|\psi_-\rangle\langle\psi_-|),
\end{equation}
whose eigenvalues $\pm I_p$ correspond to the plateaus ($|I_p|/N\simeq 0.021$s$^{-1}$) 
of the persistent-current states in Fig.~\ref{figu12} (f). Thus, the mean value of the current in the
qubit states $\langle\hat{I}\rangle_j=\langle\psi_j|\hat{I}|\psi_j\rangle$ ($j=0,1$) reads,
\begin{subequations}
\begin{equation}
\langle\hat{I}\rangle_0=I_p(1-2A^2)
\end{equation}
\begin{equation}
\langle\hat{I}\rangle_1=-I_p(1-2A^2).
\end{equation}
\label{mcurr}
\end{subequations}

\providecommand{\noopsort}[1]{}\providecommand{\singleletter}[1]{#1}%

\end{document}